\documentclass[preprint,12pt]{elsarticle}

\usepackage[utf8]{inputenc}
\usepackage[hidelinks]{hyperref}

\usepackage{xspace}
\usepackage{paralist}
\usepackage{footmisc}

\usepackage{booktabs}
\usepackage{multirow}
\usepackage{wrapfig}
\usepackage[table]{xcolor}
\usepackage{array}

\usepackage{subfigure}
\usepackage{dblfloatfix}
\usepackage{rotating}
\usepackage{tcolorbox}

\usepackage{fancyhdr}
\usepackage{amsthm}
\usepackage{amsfonts}

\usepackage[]{amsmath}
\usepackage{amssymb}
\usepackage{algorithm}
\usepackage[noend]{algpseudocode}
\algnewcommand\algorithmicinput{\textbf{Input:}}
\algnewcommand\Input{\item[\algorithmicinput]}
\algnewcommand\algorithmicoutput{\textbf{Output:}}
\algnewcommand\Output{\item[\algorithmicoutput]}
\algnewcommand{\LeftComment}[1]{\Statex \(\triangleright\) #1}
\algrenewcommand\alglinenumber[1]{\tiny #1:}

\newcommand{\Sapienz}{\textnormal{\textsc{Sapienz}}\xspace}
\newcommand{\SapienzDiv}{\textnormal{\textsc{Sapienz$^{\mathit{div}}$}}\xspace}

\newcommand{\nsgaTwo}{\mbox{NSGA-II}\xspace}

\newcommand{\ie}{\textit{i.e.}\xspace}
\newcommand{\eg}{\textit{e.g.}\xspace}
\newcommand{\cf}{\textit{cf.}\xspace}
\newcommand{\etal}{\textit{et\,al.}\xspace}

\newcommand{\aarddict}{{aarddict}\xspace}
\newcommand{\hotdeath}{{hotdeath}\xspace}
\newcommand{\kNineMail}{{k9mail}\xspace}
\newcommand{\munchLife}{{MunchLife}\xspace}
\newcommand{\passwordmanager}{{passwordmanager}\xspace}
\newcommand{\keepass}{{passwordmanager}\xspace}

\newcommand{\Atwelve}{$\hat{A}_{12}$\xspace}

\usepackage{tikz}
\usetikzlibrary{shapes}

\newcommand{\mytriangle}[2]{\tikz{\node[draw=black,fill=#1,isosceles
		triangle,isosceles triangle stretches,shape border rotate=#2,minimum
		width=0.2cm,minimum height=0.2cm,inner sep=0pt] at (0,0) {};}}
	
\newcommand{\mytriangleUp}[1]{\mytriangle{#1}{90}}
\newcommand{\mytriangleDown}[1]{\mytriangle{#1}{270}}
\newcommand{\mytriangleRight}[1]{\mytriangle{#1}{0}}

\newcommand{\betterAndSignificant}{\mytriangleUp{black}}
\newcommand{\betterAndNotSignificant}{\mytriangleUp{white}}
\newcommand{\worseAndSignificant}{\mytriangleDown{black}}
\newcommand{\worseAndNotSignificant}{\mytriangleDown{white}}
\newcommand{\equalAndNotSignificant}{\mytriangleRight{white}}
\newcommand{\equalAndSignificant}{\mytriangleRight{black}}

\newcommand{\plotwidth}{0.45}
\newcommand{\includeSubFig}[2]{
	\vspace{-.5em}
	\subfigure[\aarddict]{
		\includegraphics[width=\plotwidth\textwidth]{fig/metrics/aarddict/#2}
	}
	\subfigure[\munchLife]{
		\includegraphics[width=\plotwidth\textwidth]{fig/metrics/munchlife/#2}
	}
	\vspace{-.5em}
	\subfigure[\keepass]{
		\includegraphics[width=\plotwidth\textwidth]{fig/metrics/keepass/#2}
	}
	\vspace{-.5em}
	\subfigure[\hotdeath]{
		\includegraphics[width=\plotwidth\textwidth]{fig/metrics/hotdeath/#2}
	}
	\subfigure[\kNineMail]{
		\includegraphics[width=\plotwidth\textwidth]{fig/metrics/k9-mail/#2}
	}
	\vspace{-1em}
}

\usepackage{amssymb}

\journal{Information and Software Technology}

\begin{document}
\makeatletter
\def\ps@pprintTitle{%
	\let\@oddhead\@empty
	\let\@evenhead\@empty
	\def\@oddfoot{\parbox[t]{1\linewidth}{\footnotesize \textit{Accepted Manuscript. \@journal  \hfill September 22, 2020 \\ ~\url{https://doi.org/10.1016/j.infsof.2020.106436} \\
	License: \href{https://creativecommons.org/licenses/by-nc-nd/4.0/}{CC BY-NC-ND 4.0}	}}}%
	\let\@evenfoot\@oddfoot}
\makeatother

\begin{frontmatter}

\title{A Comprehensive Empirical Evaluation of Generating Test Suites for Mobile Applications with Diversity}

\author{Thomas Vogel}
\author{Chinh Tran}
\author{Lars Grunske}
\address{Software Engineering Group, Humboldt-Universit\"{a}t zu Berlin, Berlin, Germany {\{thomas.vogel, grunske\}@informatik.hu-berlin.de}, {mail@chinhtran.de} \vspace{-2em}}

\begin{abstract}
\textbf{Context:} In search-based software engineering we often use popular heuristics with default configurations, which typically lead to suboptimal results, or we perform experiments to identify configurations on a trial-and-error basis, which may lead to better results for a specific problem. We consider the problem of generating test suites for mobile applications (apps) and rely on \Sapienz, a state-of-the-art approach to this problem that uses a popular heuristic (NSGA-II) with a default configuration.
\textbf{Objective:} We want to achieve better results in generating test suites with \Sapienz while avoiding trial-and-error experiments to identify a more suitable configuration of \Sapienz. 
\textbf{Method:} We conducted a fitness landscape analysis of \Sapienz to analytically understand the search problem, which allowed us to make informed decisions about the heuristic and configuration of \Sapienz when developing \SapienzDiv. We comprehensively evaluated \SapienzDiv in a head-to-head comparison with \Sapienz on 34 apps.
\textbf{Results:} Analyzing the fitness landscape of \Sapienz, we observed a lack of diversity of the evolved test suites and a stagnation of the search after 25 generations. \SapienzDiv realizes mechanisms that preserve the diversity of the test suites being evolved. The evaluation showed that \SapienzDiv achieves better or at least similar test results than \Sapienz concerning coverage and the number of revealed faults. However, \SapienzDiv typically produces longer test sequences and requires more execution time than \Sapienz.
\textbf{Conclusions:} The understanding of the search problem obtained by the fitness landscape analysis helped us to find a more suitable configuration of \Sapienz without trial-and-error experiments. By promoting diversity of test suites during the search, improved or at least similar test results in terms of faults and coverage can be achieved.

\end{abstract}


\end{frontmatter}

\section{Introduction}

In search-based software engineering (SBSE) and particularly search-based testing, popular heuristics (\eg,~\cite{Mao+2016}) with best-practice configurations in terms of operators and parameters (\eg,~\cite{Fraser13A}) are often used. As this out-of-the-box usage typically leads to suboptimal results, costly trial-and-error experiments are performed to find a suitable configuration for a given problem, which leads to better results~\cite{Arcuri2013}.
To obtain better results while avoiding trial-and-error experiments, a \textit{fitness landscape analysis} can be used~\cite{Malan+Engelbrecht2013,Pitzer+Affenzeller2012}.
The goal is to analytically understand the search problem, determine difficulties of the problem, and identify suitable configurations of heuristics that can cope with these difficulties (\cf~\cite{Malan+Engelbrecht2013,Moser+2017}).

In previous work, we have investigated the search problem of test suite generation for mobile applications (apps)~\cite{2019-SSBSE}.  We relied on \Sapienz~\cite{Mao+2016} that uses a default \nsgaTwo~\cite{Deb+2002} to generate test suite for apps. \nsgaTwo has been selected by \citet[p.\,97]{Mao+2016} as it ``is a widely-used multiobjective evolutionary search algorithm, popular in SBSE research'', but without adapting it to the specific problem.
Thus, \Sapienz could likely be improved by adapting \nsgaTwo to the problem of generating test suites for apps.
For this purpose, we analyzed the fitness landscape of \Sapienz to better understand the search problem and to make informed decisions when adapting the heuristic of \Sapienz. With the adapted heuristic, we aimed for yielding better test results in terms of achieved coverage, revealed faults, and length of test sequences.
In this context, trial-and-error experiments to empirically determine a suitable configuration of \nsgaTwo was not a viable option due to the high costs of executing search-based test generation approaches for apps (\cf~\cite{2019-SSBSE,Sell+2019}).
In contrast, a fitness landscape analysis generally aims for an analytical understanding of the search problem that should help to improve the heuristic.
To the best of our knowledge, our previous work~\cite{2019-SSBSE} was the first one that analyzed the fitness landscape for the search-based generation of test suites for apps.

Our analysis of \Sapienz has focused on the global topology of the fitness landscape and evolvability, that is, how solutions (test suites) are spread in the search space and evolve over time. Thus, we are interested in the genotypic diversity of solutions being evolved, which is considered important for evolutionary search (\cf~\cite{Crepinsek2013}).
According to our analysis, \Sapienz lacks diversity of solutions so that we extended it to \SapienzDiv that integrates four diversity-promoting mechanisms with \nsgaTwo.
We have evaluated \SapienzDiv in a preliminary evaluation in our previous work~\cite{2019-SSBSE}.

In this article extending our previous paper~\cite{2019-SSBSE}, we provide a refined discussion of the required background, fitness landscape analysis of \Sapienz, \SapienzDiv approach, and related work, and particularly we present an extended empirical evaluation. This evaluation uses more apps (34\,vs.\,10), longer runs of search (40\,vs.\,10\,generations), and more repetitions of experiments (30\,vs.\,20) than the previous preliminary evaluation in order to conduct a comprehensive statistical analysis.
To further illustrate the comprehensiveness of the extended evaluation, the corresponding experiments required 562 days of execution time (\ie, the measured wall-clock time for the whole experiment) in contrast to 32 days for the previous evaluation.
The motivation to extend the search from 10 to 40 generations is that the fitness landscape analysis showed that the search of \Sapienz stagnates after 25 generations. Thus, we can evaluate whether the diversity-promoting mechanisms of \SapienzDiv will have an effect on the results when the search of \Sapienz actually stagnates.
The evaluation shows that \SapienzDiv achieves better or at least similar test results in terms of achieved coverage and revealed faults than \Sapienz, but it generates fault-revealing test sequences of similar or greater length than \Sapienz. Thus, preferring \SapienzDiv over \Sapienz for app testing could be advantageous concerning fault revelation and coverage, and disadvantageous concerning the length of test sequences that developers have to understand to debug the app and fix the fault in the app.
Such an evaluation and discussion were not possible in our previous work due to the preliminary evaluation limiting the search to 10 generations.
Therefore, the contributions of this article are the refined discussion of our work and particularly the comprehensive evaluation of \SapienzDiv aligned with the results of the fitness landscape analysis of \Sapienz.

The rest of the article is structured as follows.
We discuss the background of our work in Section~\ref{sec:background},
the descriptive study analyzing the fitness landscape of \Sapienz in Section~\ref{sec:fla-sapienz},
\SapienzDiv in Section~\ref{sec:sapienzdiv},
the comprehensive evaluation in Section~\ref{sec:evaluation},
threats to validity in Section~\ref{sec:threats-to-validity},
and related work in Section~\ref{sec:related-work}.
Finally, we conclude the article with Section~\ref{sec:conclusion}.

\section{Background}
\label{sec:background}

In this section, we introduce \Sapienz and the basic idea of a fitness landscape analysis to obtain an analytical understanding of a search problem.

\subsection{Test Suite Generation with \Sapienz}
\label{sec:sapienz}

\Sapienz is a multi-objective search-based testing approach~\cite{Mao+2016}. Using \nsgaTwo, it automatically generates test suites for system-level end-to-end testing of Android apps.
A test suite~$t$ consists of $m$ ordered test cases
$\left\langle s_1, s_2,...,s_m \right\rangle$,
each of which is a sequence of up to $n$~GUI-level events
$\left\langle e_1, e_2,...,e_n \right\rangle$
that exercise the app under test. Examples of such events are clicks and gestures.

To evolve a population of such test suites, a whole test suite variation operator is used by \Sapienz. It consists of a uniform set element crossover among test suites applied with probability $p$ to achieve inter-individual variation, and a complex mutation operator changing single test suites to achieve intra-individual variation. This mutation operator shuffles the order of test cases within a test suite, subsequently applies a single-point crossover on two neighboring test cases of the suite with probability $q$, and finally shuffles the order of events within each test case of the suite with probability $q$.
For the selection of test suites, \Sapienz uses the select operator of \nsgaTwo. This select operator prefers individuals with a better fitness (smaller non-domination rank) and, when the fitness is equal, individuals with a greater diversity in the objective space (greater crowding distance).
For further details of the \Sapienz operators, we refer to~\cite{Mao+2016}.

The evolution of test suites is guided by three objectives:
\begin{inparaenum}[(i)]
	\item maximize fault revelation,
	\item maximize coverage, and
	\item minimize test sequence length.
\end{inparaenum}
Thus, the generated test suites should find many faults and ideally exercise the whole app while the lengths of the test sequences/cases of a suite should be minimized. Shorter test sequences are easier to be understood and reproduced by developers to identify and fix the faults revealed by them.
Having no oracle, \Sapienz considers a crash of the app caused by a test sequence as a fault.
Coverage is measured at the code level (statement coverage) if the source code of the app is available, otherwise at the Android activity level (skin coverage).
Given these objectives, the fitness function is the triple of the number of crashes found, achieved coverage, and sequence length.
Thus, the search of \Sapienz results in a Pareto front of test suites that are the best trade-offs with respect to the three objectives.

To evaluate the fitness of a test suite, \Sapienz executes the suite on the app under test deployed on an Android device or emulator. In this context, \Sapienz supports parallel evaluation of multiple test suites to improve the efficiency of the search by using several devices or emulators concurrently.

\subsection{Fitness Landscape Analysis}
\label{sec:fla}

According to \citet{Culberson1998} an evolutionary algorithm is a black box as there is no clear understanding of how its operators and parameters settings interact. Therefore, ``[t]he researcher trying to solve a problem is then placed in the unfortunate position of having to find a representation, operators, and parameter settings to make a poorly understood system solve a poorly understood problem. In many cases he might be better served by concentrating on the problem itself''~\cite[p.\,125]{Culberson1998}.
In this context, a \textit{fitness landscape analysis} is one approach to obtain a better understanding of the search problem before making any decision about the algorithm~\cite{Malan+Engelbrecht2013}.
The insights gained from the analysis should allow researchers and practitioners to make informed decisions about the algorithm to solve the problem.

According to \citet{Stadler2002}, a fitness landscape is defined by three elements:
\begin{enumerate}[(1)]\itemsep-0.4em
	\item A search space as a set $X$ of potential solutions.
	\item A fitness function $f_k : X \rightarrow {\rm I\!R}$ for each of the $k$ objectives.
	\item A neighborhood relation $N : X \rightarrow 2^{X}$ that associates neighbor solutions to each solution.
\end{enumerate}
The neighborhood relation is typically based on the search operators of the used algorithm that perform small changes of a solution to obtain a new solution, for instance, the mutation operator of a genetic algorithm. In this case, two solutions are neighbors if by applying the mutation operator to one solution, the other solution can be reached (\cf~\cite{Moser+2017}).
According to \citet[p.\,409]{Moser+2017}, it is common that the neighborhood relation is based on such local search operators since the goal is to describe fitness ``landscapes where the distance between a solution and its neighbour is shorter than that between the solution and the neighbour's neighbour''.
Accordingly, \citet{Stadler2002} associates the notions of nearness, distance, or accessibility to the neighborhood relation.
Thus, it is possible to quantify the neighborhood relation using a distance metric that is based on the local search operators of the genetic algorithm being used.
Relying on a search operator of an algorithm for the neighborhood relation, each operator describes an individual fitness landscape, which is called ``one operator, one landscape'' by \citet{Jones95PhD}. Therefore, the fitness landscape can be seen as a description of the (1)~search space as explored by the genetic algorithm that uses the (2)~fitness functions for guiding the search and whose operators determine the (3)~neighborhood~relation.

Given these three elements defining a fitness landscape, various metrics have been proposed to analyze the landscape~\cite{Malan+Engelbrecht2013,Pitzer+Affenzeller2012}. They characterize the landscape, for instance, in terms of
the global topology (\ie, the distributions of solutions and fitness within the landscape),
local structure (\ie, the number and distribution of local optima, the shape of landscape areas in terms of ruggedness and smoothness, or plateaus as connected areas of equal fitness), and
evolvability (\ie, the ability to produce fitter solutions denoting a progress of the search).
The goal of analyzing the landscape is to determine difficulties of a search problem and identify suitable configurations of search algorithms that can cope with these difficulties (\cf~\cite{Malan+Engelbrecht2013,Moser+2017}).
For this purpose, the fitness landscape analysis opens the black box of a genetic algorithm and investigates how the algorithm explores the search space.

\section{Fitness Landscape Analysis of \Sapienz}
\label{sec:fla-sapienz}

In this section, we discuss our descriptive study analyzing the fitness landscape of \Sapienz. For this purpose, we define the fitness landscape of \Sapienz in Section~\ref{sec:fl-sapienz} and present the design, experimental setup, and results of analyzing the fitness landscape of \Sapienz in Sections~\ref{sec:fla-design} and~\ref{sec:fla-results}. Finally, we discuss these results in Section~\ref{sec:fla-discussion}.

\subsection{Fitness Landscape of \Sapienz}
\label{sec:fl-sapienz}

First, we define the fitness landscape of \Sapienz in terms of the three elements of a fitness landscape (\cf~Section~\ref{sec:fla}):
\begin{enumerate}[(1)]\itemsep-0.4em
	\item The search space of \Sapienz is given by all possible test suites $t$ according to the representation of test suites in \Sapienz (Section~\ref{sec:sapienz}).
	\item The fitness function is given by the triple of the number of crashes found, achieved coverage, and test sequence length (Section~\ref{sec:sapienz}).
	\item As the neighborhood relation we define the metric $dist(t_1, t_2)$ that computes the distance between two test suites $t_1$ and $t_2$ at the genotypic~level.
\end{enumerate}

The distance metric $dist(t_1, t_2)$ (Algorithm~\ref{alg:distance}) has as input two parameters, the test suites $t_1$ and $t_2$ as produced by \Sapienz, whereas the test suite size $m$ and the maximal test sequence length $n$ are  determined by the static configuration of \Sapienz. The metric computes the genotypic distance between $t_1$ and $t_2$, which is the output of the algorithm, as follows.
The distance between two test suites $t_1$ and $t_2$ is the sum of the distances between their ordered test sequences, which is obtained by comparing all sequences $s^{t1}_i$ of $t_1$ and $s^{t2}_i$ of $t_2$ by index~$i$ (lines~\ref{alg:distance:l2}--\ref{alg:distance:l4}). The distance between two test sequences $s^{t1}_i$ and $s^{t2}_i$ is the difference of their lengths (line~\ref{alg:distance:l5}) increased by~$1$ for each different event at index~$j$ (lines~\ref{alg:distance:l6}--\ref{alg:distance:l9}). Thus, the distance is based on the differences of ordered events between the ordered test sequences of two test suites.

\begin{algorithm}[t]
	\caption{$dist(t_1, t_2)$: compute distance between two test suites $t_1$ and~$t_2$.}
	\label{alg:distance}
	\scriptsize
	\begin{algorithmic}[1]
		\Input Test suites $t_{1}$ and $t_{2}$, test suite size $m$, maximal test sequence length $n$
		\Output Distance between $t_{1}$ and $t_{2}$
		\State \texttt{distance $\leftarrow 0$;}
		\For{\texttt{i $\leftarrow$ 0 to $m$}} \Comment{iterate over all $m$ test sequences} \label{alg:distance:l2}
		\State \texttt{$s^{t1}_i \leftarrow t_{1}[i]$;} \Comment{$i^{th}$ test sequence of test suite $t_1$}
		\State \texttt{$s^{t2}_i \leftarrow t_{2}[i]$;} \Comment{$i^{th}$ test sequence of test suite $t_2$} \label{alg:distance:l4}
		\State \texttt{distance $\leftarrow$ distance + abs(|$s^{t1}_i$| - |$s^{t2}_i$|);} \Comment{length difference as\,distance} \label{alg:distance:l5}
		\For{\texttt{j $\leftarrow$ 0 to $n$}}  \Comment{iterate over all $n$ events} \label{alg:distance:l6}
		\If{\texttt{|$s^{t1}_i$| $\leq j$ \textbf{or} |$s^{t2}_i$| $\leq j$}} break;
		\EndIf
		\If{\texttt{$s^{t1}_i$[j] $\neq$ $s^{t2}_i$[j]}} \Comment{event comparison by index $j$}
		\State \texttt{distance $\leftarrow$ distance + 1;} \Comment{events differ at index $j$ in both seqs.} \label{alg:distance:l9}
		\EndIf
		\EndFor
		\EndFor
		\State \textbf{return} \texttt{distance};
	\end{algorithmic}
\end{algorithm}

Since it is common that the neighborhood relation is based on the search operators of the genetic algorithm being used that make small changes to solutions (\cf~Section~\ref{sec:fla}), we developed this distance metric according to the mutation operator of \Sapienz that performs small changes on individual test suites. The smallest changes performed by mutation are the shuffling of the order of test sequences within a test suite, and the order of events within a test sequence.
Consequently, to align the neighborhood relation for the fitness landscape to the search operators of \Sapienz, we defined a corresponding distance metric rather than using a standard edit distance.

\subsection{Design and Experimental Setup of the Fitness Landscape Analysis}
\label{sec:fla-design}

\newcommand{\numberOfMetrics}{11\xspace}
\newcommand{\ppos}{{\textit{ppos}}\xspace}
\newcommand{\hv}{{\textit{hv}}\xspace}
\newcommand{\diam}{{\textit{diam}}\xspace} 
\newcommand{\maxdiam}{{\textit{maxdiam}}\xspace}
\newcommand{\mindiam}{{\textit{mindiam}}\xspace}
\newcommand{\avgdiam}{{\textit{avgdiam}}\xspace}
\newcommand{\reldiam}{{\textit{reldiam}}\xspace}
\newcommand{\pconnec}{{\textit{pconnec}}\xspace}
\newcommand{\nconnec}{{\textit{nconnec}}\xspace}
\newcommand{\kconnec}{{\textit{kconnec}}\xspace}
\newcommand{\lconnec}{{\textit{lconnec}}\xspace}
\newcommand{\hvconnec}{{\textit{hvconnec}}\xspace}

Among the global topology, local structure, and evolvability that can be investigated by a fitness landscape analysis (Section~\ref{sec:fla}), we are interested in the global topology, that is, how solutions (test suites) are spread in the search space during the search process of \Sapienz. Thus, we investigate the genotypic diversity of the solutions being evolved since it is known that diversity has an influence on the performance of genetic algorithms (\cf~\cite{Crepinsek2013}).
For instance, \citet{PanichellaOPL15} have shown in the context of regression testing that the selection of test cases based on genetic algorithms could be improved by diversifying the test cases during the search. However, the diversity of solutions in the context of generating test suites for apps has not been investigated yet by a fitness landscape analysis.
Besides the diversity of the test suites during the search, we investigate the evolvability, that is, the progress of the search and how the test suites evolve over time.

To analyze the diversity and evolution of test suites in \Sapienz, we selected \textit{\numberOfMetrics different metrics} from literature since ``[n]o single measure or description can possibly characterize any high-dimensional heterogeneous search space''~\cite[p.\,31]{SmithHLO02} or the difficulty of the related search problems~\cite{Malan+Engelbrecht2013,Pitzer+Affenzeller2012}. This view holds for \Sapienz and the generation of test suites for apps representing such a complex search space and problem.
These $\numberOfMetrics$ metrics characterize
the evolvability of the search and
the diversity of the whole population and Pareto-optimal solutions.
We implemented them in \Sapienz and compute them after every generation of the search process so that we can analyze their development over time, and thus, the progress of the search and how the solutions and their diversity evolve.
We will discuss each metric when presenting the results of the fitness landscape analysis in the next section.

\begin{table}[t]
	\centering
	\caption{Apps and their test results~\cite{Mao+2016} we selected for the fitness landscape analysis.}
	\label{tab:fla-apps}
	\resizebox*{1\textwidth}{!}{
		\begin{tabular}{lclccc} \toprule
			\textbf{Name} & \textbf{Version} & \textbf{Category} & \textbf{Coverage (\%)} & \textbf{\#Crashes} & \textbf{Seq. Length} \\ \midrule
			\aarddict & 1.4.1 & Books \& Reference & 18 & 0 & - \\
			\keepass & 1.9.8 & Tools & 16 & 0 & - \\
			\munchLife & 1.4.2 & Entertainment & 76 & 0 & - \\
			\hotdeath & 1.0.7 & Card & 79 & 3 & 152 \\
			\kNineMail & 5.207 & Communications & 7 & 1 & 238 \\
			\bottomrule
		\end{tabular}
	}
\vspace{-1em}
\end{table}

For the fitness landscape analysis of \Sapienz, we set up an experiment.
Due to the high execution costs of \Sapienz, we selected five apps (Table~\ref{tab:fla-apps}) from the 68 F-Droid apps benchmark~\cite{Choudhary+2015} used to evaluate \Sapienz~\cite{Mao+2016}.
Particularly, we selected
\textit{\aarddict},
\textit{\passwordmanager},
and
\textit{\munchLife}
since \Sapienz did not find any fault for them,
and \textit{\hotdeath}
and \textit{\kNineMail}\footnote{We used version 5.207 of \kNineMail instead of 3.512 as in the 68 F-Droid apps benchmark~\cite{Choudhary+2015} due to unavailability of the respective version at the time of study.
The other four apps are of the same version as in the benchmark.
}
for which \Sapienz did find faults~\cite{Mao+2016}. 
Moreover, the individual apps are from different app categories and \Sapienz achieved different levels of statement coverage and generated fault-revealing test sequences of different lengths for them (Table~\ref{tab:fla-apps}).
Thus, we consider apps with different test results to obtain potentially different fitness landscape features that may present difficulties to the \Sapienz search.

To conduct the experiment, we execute \Sapienz extended with the $\numberOfMetrics$ metrics on each of the five selected apps, repeat each execution five times, and report for each app the time series (over the generations) of the metric values averaged over the repetitions.\footnote{All experiments were run on a single 4.0 GHz quad-core PC with 16\,GB RAM, using five Android emulators (KitKat 4.4.2, API level 19) in parallel to test one app.}
We use the default configuration of \Sapienz~\cite{Mao+2016} listed in Table~\ref{tab:sapienz-config}. The crossover and mutation rates are set to 0.7 and 0.3 respectively. The population size and offspring size are 50. An individual (test suite) contains five test sequences, each constrained to 20--500 events. However, we set the stopping criterion of the search to 40 generations to obtain longer runs than in~\cite{Mao+2016}.

\begin{table}[t]
	\small
	\centering
	\caption{\Sapienz configuration.}
	\label{tab:sapienz-config}
	\resizebox*{.8\textwidth}{!}{
		\begin{tabular}{lr}
			\toprule
			\textbf{Parameter} & \textbf{Value} \\
			\midrule
			Crossover rate & 0.7 \\
			Mutation rate & 0.3 \\
			Population size & 50 \\
			Offspring size & 50 \\
			Test suite size (number of test cases per test suite) & 5 \\
			Minimum length of a test sequence (number of events) & 20 \\
			Maximum length of a test sequence (number of events) & 500 \\
			Number of generations (stopping criterion of the search) & 40 \\
			\bottomrule
		\end{tabular}
	}
\vspace{-1em}
\end{table}

\subsection{Results of the Fitness Landscape Analysis}
\label{sec:fla-results}

The results of our descriptive study provide an analysis of the fitness landscape of \Sapienz with respect to the
evolvability, that is, the progress of the search, and
global topology, that is, the diversity and spread of solutions (test suites) in the search space.
In the following, we present these results by discussing the $\numberOfMetrics$ metrics we implemented in \Sapienz and the time series we obtained for these metrics by executing \Sapienz on each of the five selected apps.
The discussion is grouped by the purpose of the metrics as they characterize either
(1)~the evolvability of the search,
(2)~the diversity of the whole population, or
(3)~the diversity of Pareto-optimal solutions.

\subsubsection{Metrics for the Evolvability of the Search}
The following two metrics characterize the evolvability of the search by looking at the evolution of the Pareto-optimal solutions during the search.

\paragraph{Proportion of Pareto-optimal solutions (\ppos)}
For a population $P$, \ppos is the number of Pareto-optimal solutions $P_{opt}$ (\ie,  the non-dominated solutions of $P$) divided by the population size: $ppos(P) = \frac{|P_{opt}|}{|P|}$.
A high and especially strongly increasing \textit{ppos} may indicate a problem of \textit{dominance resistance}, that is, the search cannot produce new solutions by crossover and mutation that dominate the current, poorly performing but locally non-dominated solutions. Due to this missing selection pressure, the search based on Pareto dominance may stagnate, potentially far away from the true Pareto front~\cite{Purshouse+Fleming2007}. 
A moderately increasing \ppos may indicate a successful search.

\begin{figure}[t!]
	\centering
	\includeSubFig{\ppos}{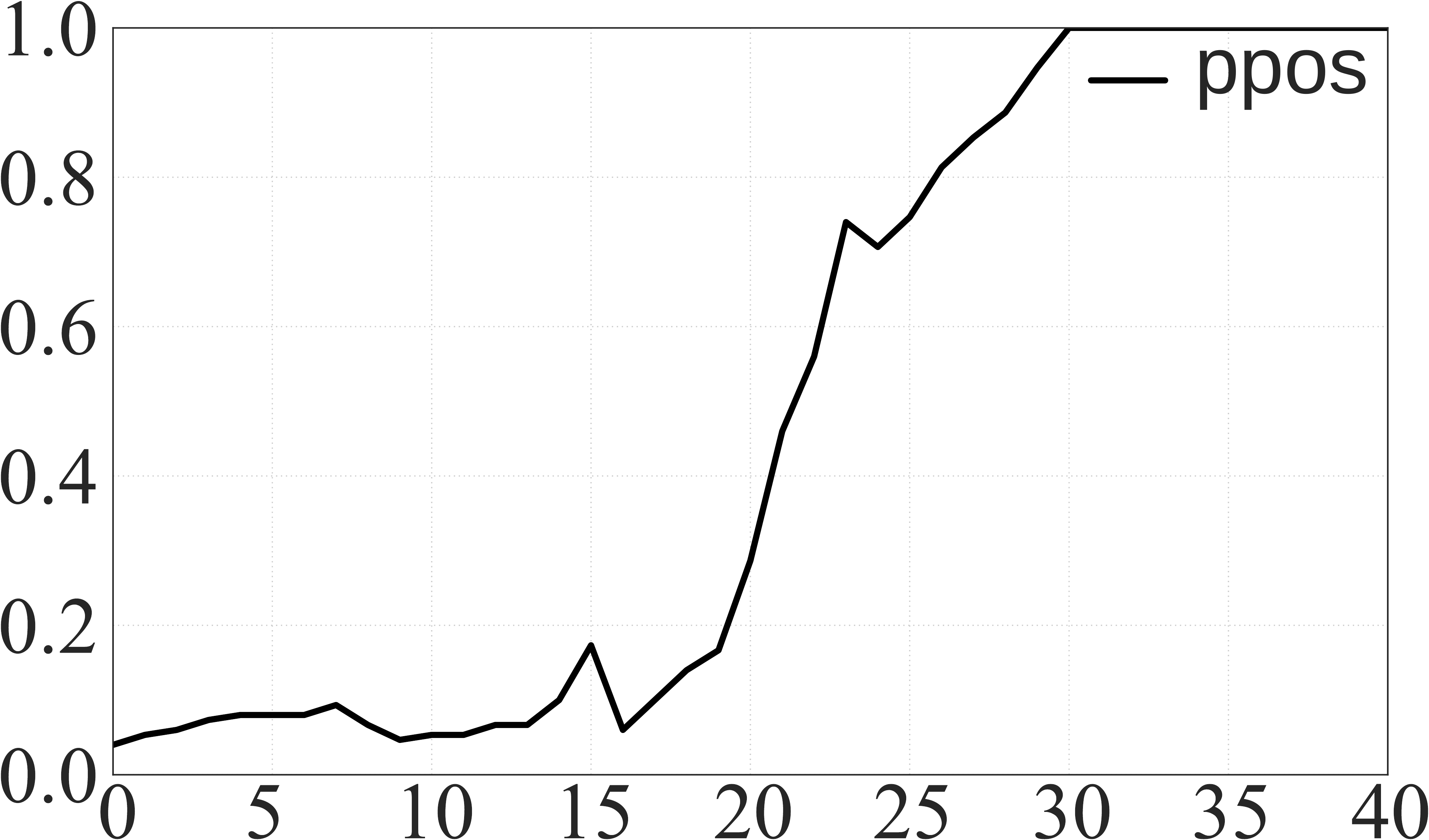}
	\caption{Proportion of Pareto-optimal solutions (\ppos) over the 40 generations of search.}
	\label{fig:ppos}
	\vspace{-1em}
\end{figure}

For \Sapienz and all apps (see~Fig.~\ref{fig:ppos} plotting \ppos averaged over the five runs along the 40 generations of the search for each of the five apps), \ppos slightly fluctuates since a single new solution from the current generation can potentially dominate multiple solutions that have been non-dominated in the previous generation. In this case, \ppos decreases. In contrast, \ppos increases if more new solutions are found than such previously non-dominated solutions that are now dominated by the new solutions, or if new solutions just extend the Pareto front without dominating any other non-dominated solution of the previous generation.
At the beginning of the search, \ppos is low ($<$0.1), shows no improvement in the first 15--20 generations, and then increases for all apps except of \textit{\passwordmanager}.
Thus, the search seems to progress while the enormously increasing \ppos for \munchLife and \hotdeath after around 20 and 25 generations respectively might indicate a stagnation of the search.

\paragraph{Hypervolume (\hv)}
To further investigate the search progress, we compute the \hv after each generation. The \hv is the volume in the objective space covered by the Pareto-optimal solutions~\cite{Li+Yao2019,Wang+2016}. According to \citet{Li+Yao2019}, it is the most popular quality indicator in optimization and used to compare different Pareto fronts or generally, two sets of solutions. If one set is better than the other with respect to the objectives of the search, then the \hv of the better set is higher than the \hv of the other set.
Thus, within a search process, an increasing \hv indicates that the search is able to find improved solutions, otherwise the \hv and search stagnate.

Based on the objectives of \Sapienz (\ie, maximize the number of crashes, maximize the achieved coverage, and minimize the length of test sequences), we choose the nadir point as the reference point for the \hv, that is, $0$~crashes, $0$~coverage, and a test sequence length of $500$ (we configured \Sapienz to generate test sequences that consists of at most 500 events (\cf~Section~\ref{sec:fla-design})).

\begin{figure}
	\centering
	\includeSubFig{\hv}{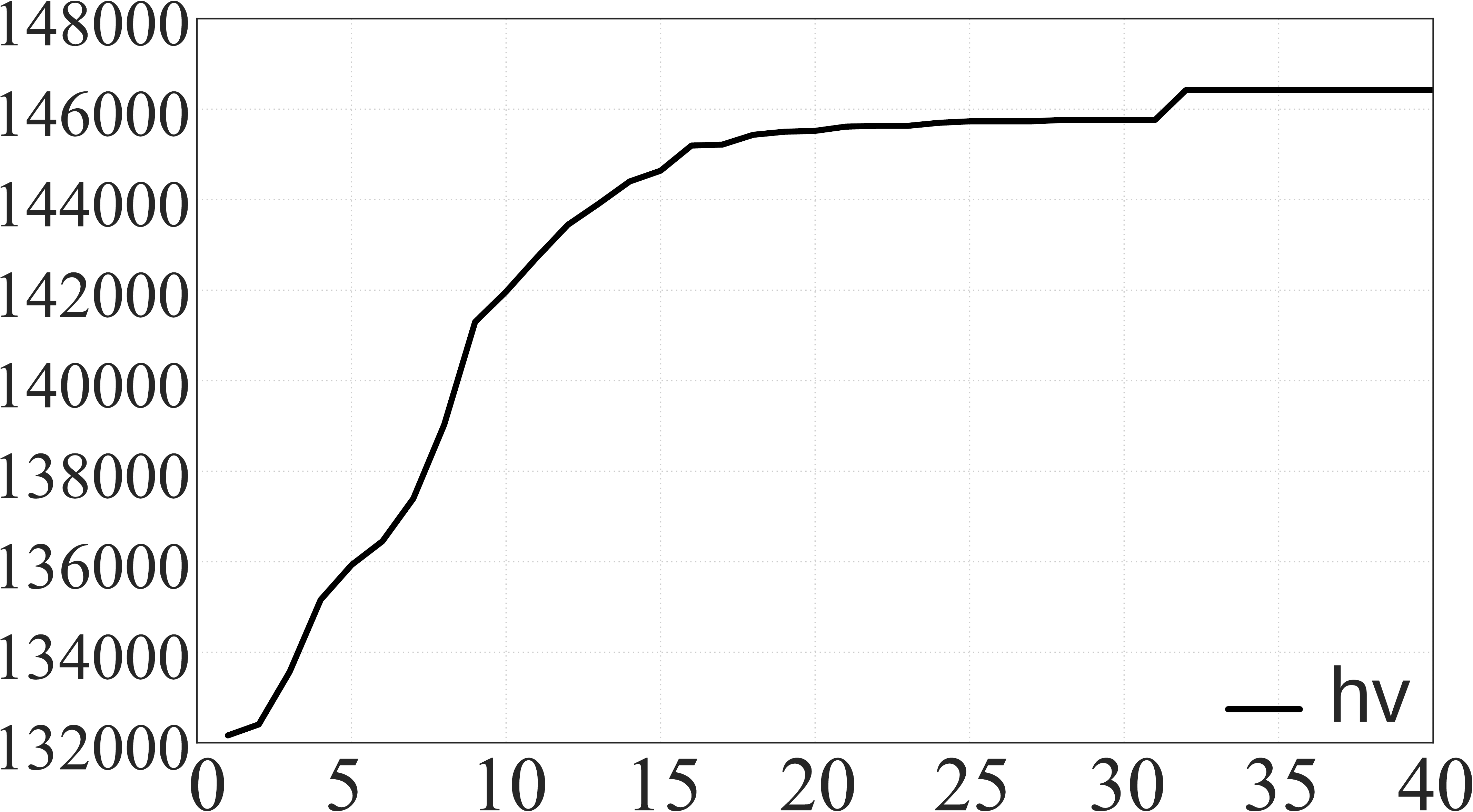}
	\caption{Hypervolume (\hv) over the 40 generations of search.}
	\label{fig:hv}
	\vspace{-1em}
\end{figure}

In Fig.~\ref{fig:hv}, the evolution of the \hv over time rather than the absolute numbers are relevant to analyze the search progress of \Sapienz.
While the \hv increases during the first 25 generations, it stagnates afterwards for all apps; for \kNineMail already after 5 generations. Thus, the search of \Sapienz is able to find improved solutions but it stagnates after 25 generations. 
For \aarddict, \munchLife, and \hotdeath the \hv stagnates after the \ppos drastically increases (\cf~Fig.~\ref{fig:ppos}), further indicating a stagnation of the search after 20--25 generations.

\subsubsection{Metrics for the Diversity of the Population}\label{sec:fla:divpop}
The following four metrics characterize the diversity of the whole population, that is, of \textit{all} individuals in each generation of the search.

\paragraph{Maximum, Average, and Minimum Population diameter (\maxdiam, \avgdiam, \mindiam)}
These metrics measure the spread of all population members in the search space using a distance metric for individuals. In our case, we use the metric $dist(t_1, t_2)$ defined by Algorithm~\ref{alg:distance} computing the distance between any two test suites $t_1$ and $t_2$.
The maximum population diameter computes the largest distance between any two individuals of the population~$P$: $\maxdiam(P) = \max_{x_i, x_j \in P} dist(x_i, x_j)$ \cite{Bachelet99,Olorunda+Engelbrecht2008}. Thus, \maxdiam shows the absolute spread of $P$ in the search space.
To respect outliers, we can compute the average population diameter as the average of all pairwise distances between all individuals~\cite{Bachelet99}:
\begin{equation}
	\avgdiam(P) = \frac{\sum_{i=0}^{|P|}\sum_{j=0,j \neq i}^{|P|}{dist(x_i, x_j)}}{ |P|(|P|-1)}
	\label{eq:avgdiam}
\end{equation}
Additionally, we compute the minimum population diameter to see how close individuals are in the search space: $\mindiam(P) = \min_{x_i, x_j \in P} dist(x_i, x_j)$. Using \mindiam, we can test whether individuals are even identical and thus represent duplicate solutions in a population. In this case, \mindiam is $0$.

\begin{figure}[t!]
	\vspace{-1em}
	\centering
	\includeSubFig{\diam}{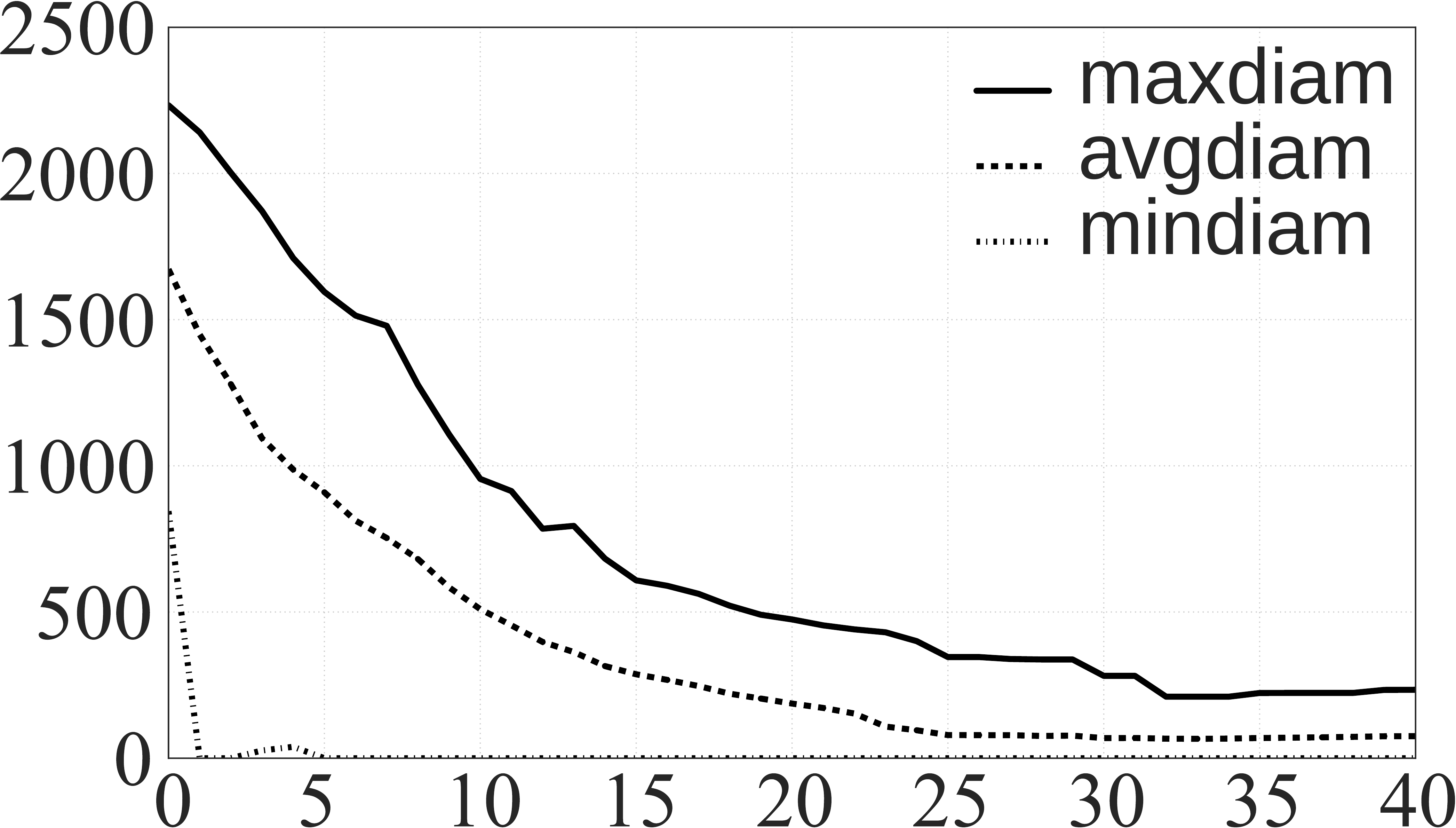}
	\caption{Maximum, average, and minimum population diameters (\maxdiam, \avgdiam, \mindiam) over the 40 generations of search.}
	\label{fig:diam}
\end{figure}

Concerning each plot in Fig.~\ref{fig:diam}, the upper, middle, and lower curve are respectively \maxdiam, \avgdiam, and \mindiam.
Since all of these three metrics rely on our distance metric $dist(t_1, t_2)$, their ranges are determined by the distance metric that ranges from $0$ (the two solutions $t_1$ and $t_2$ are identical) to $2500$ (the two solutions $t_1$ and $t_2$ differ as much as possible, that is, they differ in all events---up to 500 events for a test sequence---for all of their five individual test sequences).
For each curve, we see a clear trend that the metrics decrease over time, which is typical for genetic algorithms due to the crossover.
However, the metrics drastically decrease for \Sapienz in the first 25 generations. The \avgdiam decreases from $>$$1500$ to eventually $<$$200$ for each app. The \maxdiam decreases similarly from around $2250$ to $\approx$$250$ but stays higher for \hotdeath ($>$$500$) and \kNineMail ($\approx$$1000$), which suggests that there are a few outliers in these two apps.
The development of the \avgdiam and \maxdiam indicates that all individuals are continuously getting closer to each other in the search space, thus becoming more similar.
The population even contains identical solutions as indicated by \mindiam reaching~$0$.

\paragraph{Relative population diameter (\reldiam)}
Besides the population diameter metrics discussed previously, \citet{Bachelet99} further proposes the relative population diameter (\reldiam), which is the average population diameter (\avgdiam) in proportion to the largest possible distance $d$: $\reldiam(P) = \frac{avgdiam(P)}{d}$.
This metric is indicative of the concentration of the population in the search space. Especially, a small \reldiam indicates that the population members are grouped together in a region of the space~\cite{Bachelet99}.

\begin{figure}[b!]
	\centering
	\includeSubFig{\reldiam}{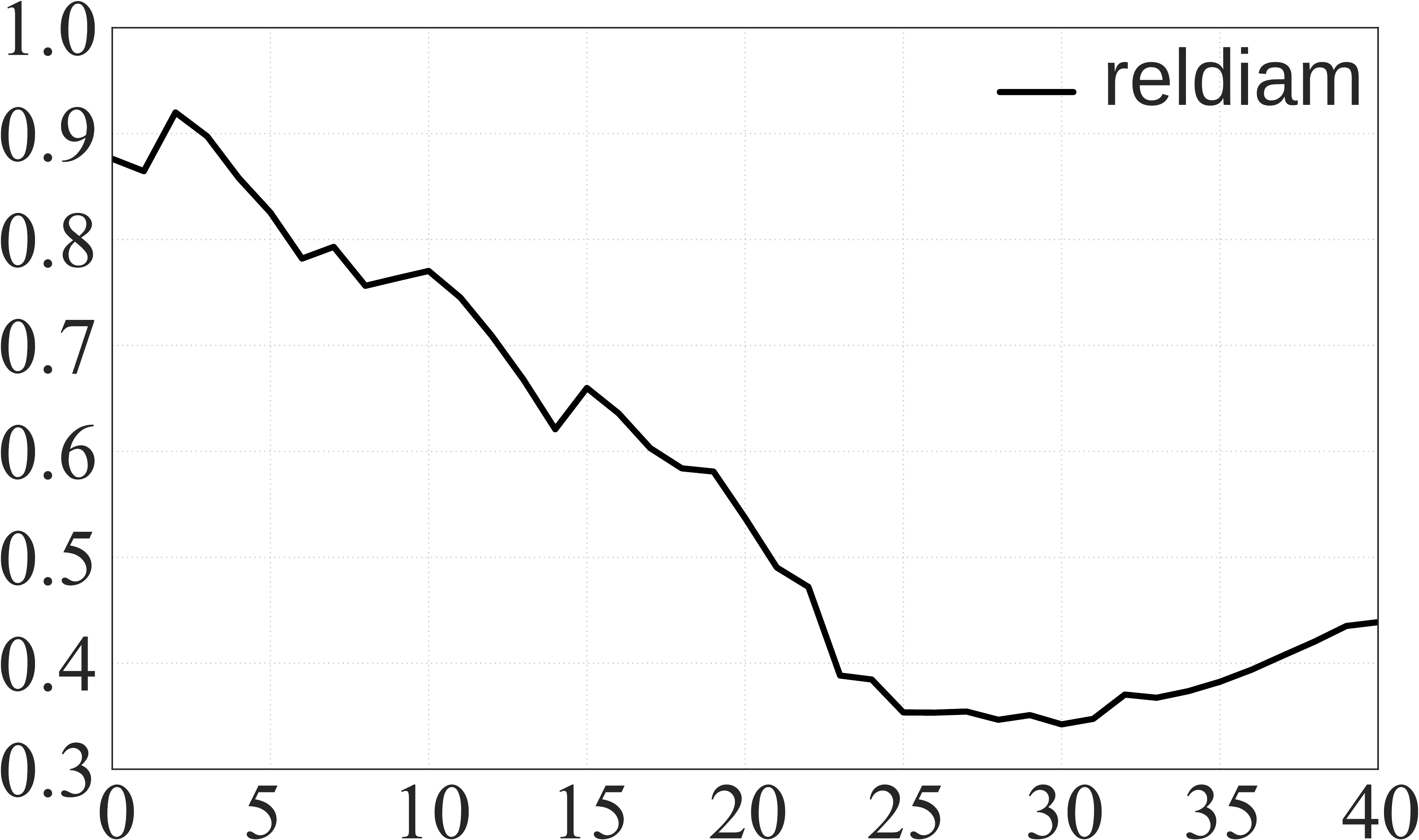}
	\caption{Relative population diameter (\reldiam) over the 40 generations of search.}
	\label{fig:reldiam}
\end{figure}

For \Sapienz, the largest possible distance $d$ between two test suites is 2500, in which case they differ in all events (up to 500 for a test sequence) for all of their five individual test sequences (\cf~previous discussion of the diameter metrics). For $d = 2500$ and all apps (\cf~Fig.~\ref{fig:reldiam}), \reldiam starts at a high level of around 0.9 indicating that the solutions are spread in the search space. Then, it decreases in the first 25 generations to around 0.4 for \aarddict, \munchLife, and \passwordmanager, and below 0.3 for \hotdeath and \kNineMail indicating a grouping of the solutions in a region of the search space. Thus, we observe a continuous concentration process of the population in the search space in the first 25 generations, after which the concentration stays high for \aarddict, \hotdeath, and \kNineMail (\reldiam is relatively constant), or slightly decreases for \munchLife and \passwordmanager (\reldiam slightly~increases).

\subsubsection{Metrics for the Diversity of the Pareto-Optimal Solutions}
\label{sec:fla:connectedness}

The following metrics analyze the \textit{connectedness} and thus, clusters of Pareto-optimal solutions in the search space~\cite{Isermann77,Paquete09}.
For this purpose, we consider a graph in which Pareto-optimal solutions are vertices.
The edges connecting the vertices depend on the adjacency of Pareto-optimal solutions defined by a neighborhood operator.
Particularly, the edges are labeled with weights $\delta$, which are the number of moves a neighborhood operator has to make to reach one vertex from another~\cite{Paquete09}. This results in a graph of fully connected Pareto-optimal solutions.
Introducing a limit $k$ on $\delta$ and removing the edges whose weights $\delta$ are larger than $k$ leads to varying sizes of connected components (clusters) in the graph.
This graph can be analyzed by metrics to characterize the Pareto-optimal solutions in the search space~\cite{Paquete09,Liefooghe14}.

In our case, the weights $\delta$ are determined by the distance metric $dist(t_1, t_2)$ for two test suites $t_1$ and $t_2$ (\cf~Algorithm~\ref{alg:distance}) that is based on the mutation operator of \Sapienz (\cf~Section~\ref{sec:fl-sapienz}).
We determined $k$ experimentally to be $300$ investigating values of $400$, $300$, $200$, and $100$. While a high value results in a single cluster of Pareto-optimal solutions, a low value results in a high number of singletons (\ie, many clusters each containing a single Pareto-optimal solution). Thus, two test suites (vertices) are connected (neighbors) in the graph if they differ in less than $300$ events across their five individual test sequences as computed by the distance metric $dist(t_1, t_2)$. 
Based on such a graph of Pareto-optimal solutions for \Sapienz, we compute in each generation of the search the following five metrics to analyze the Pareto-optimal solutions, their connectedness, and the corresponding evolution over~time.

\paragraph{Proportion of Pareto-optimal solutions in clusters (\pconnec)}
This metric divides the number of vertices (Pareto-optimal solutions) that are members of clusters (excluding singletons) by the total number of vertices in the graph~\cite{Paquete09}. A high \pconnec indicates a grouping of the Pareto-optimal solutions in one or more areas of the search space while a low value indicates a spread of these solutions across the search space.

\begin{figure}[t!]
	\centering
	\includeSubFig{\pconnec}{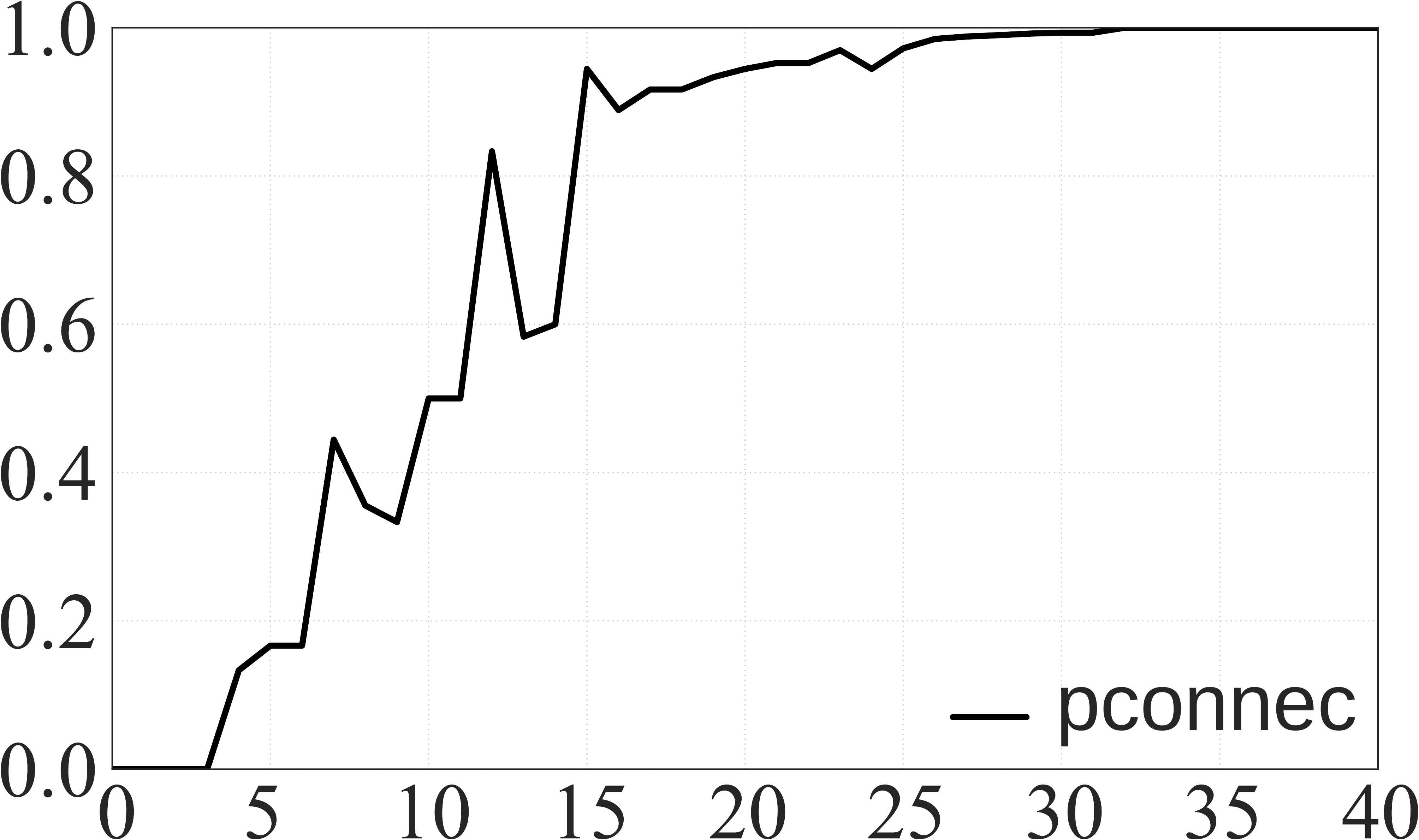}
	\caption{Proportion of Pareto-optimal solutions in clusters (\pconnec) over the 40 generations of search.}
	\label{fig:pconnec}
	\vspace{-1em}
\end{figure}

As shown in Fig.~\ref{fig:pconnec}, \pconnec is relatively low during the first generations before it increases for all apps. For \munchLife, \passwordmanager, and \hotdeath, \pconnec reaches 1 meaning that all Pareto-optimal solutions are located in (one or more) clusters, while it converges around 0.7 and 0.8 for \aarddict and \kNineMail respectively. Overall, this indicates that all or at least most of the Pareto-optimal solutions are grouped in one or more areas of the search space.

\paragraph{Number of clusters (\nconnec)}
We further analyze in how many areas of the~search space (clusters) the Pareto-optimal solutions are grouped.
Thus, \nconnec counts the number of clusters in the graph~\cite{Paquete09,Liefooghe14}. A high \nconnec indicates that the Pareto-optimal solutions are spread in many areas, whereas a low \nconnec indicates that the Pareto-optimal solutions are spread in few areas of the search space.

\begin{figure}[t!]
	\centering
	\includeSubFig{\nconnec}{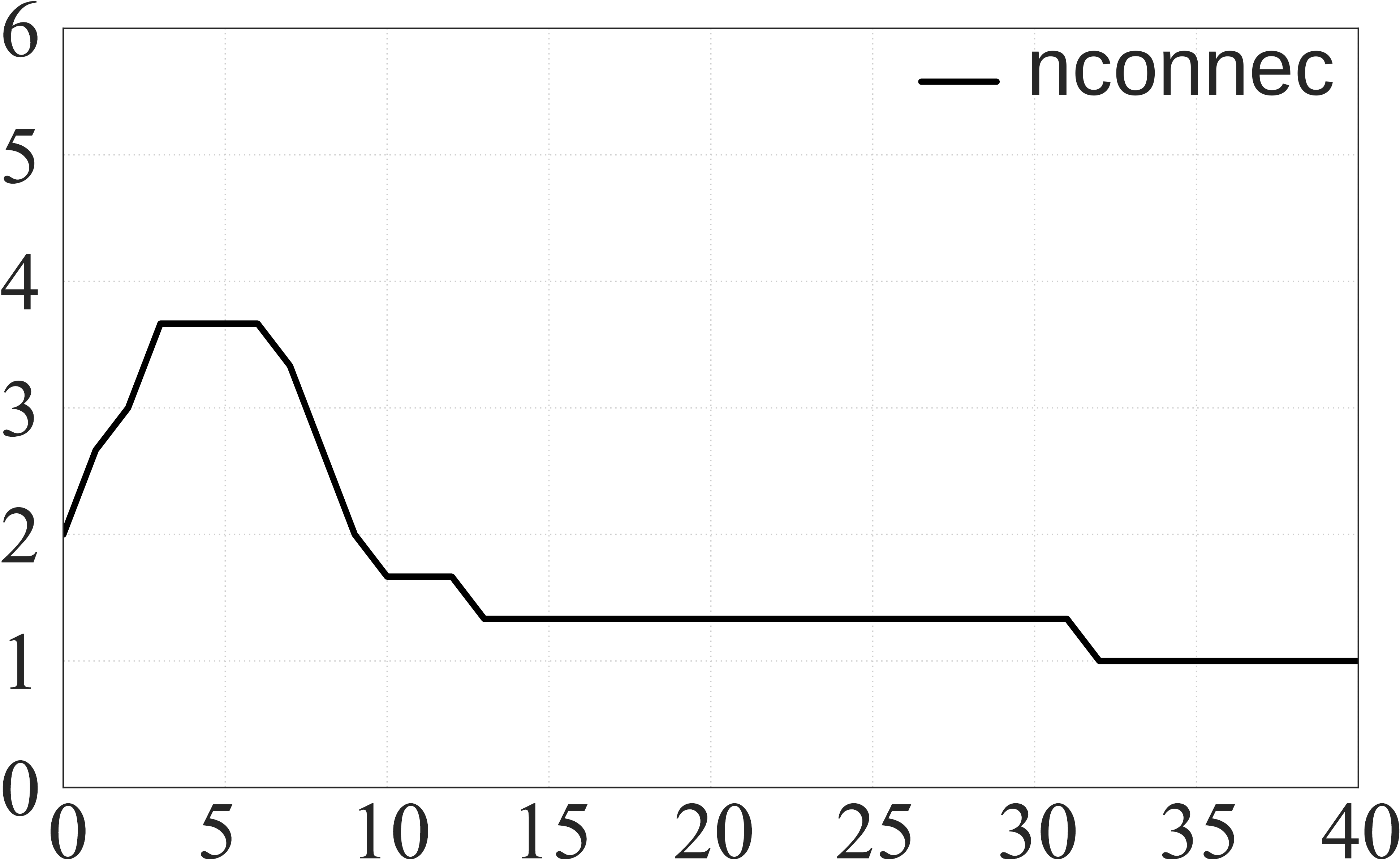}
	\caption{Number of clusters (\nconnec) over the 40 generations of search.}
	\label{fig:nconnec}
	\vspace{-1em}
\end{figure}

Fig.~\ref{fig:nconnec} plots \nconnec for \Sapienz and all apps. The y-axis of each plot denoting \nconnec ranges from $0$ to $6$.
Initially, the Pareto-optimal solutions are distributed in 2--4 clusters, then grouped in $1$ cluster. A noticeable exception is \kNineMail for which there always exist more than $3$ clusters. Except for \kNineMail, this indicates that the Pareto-optimal solutions are grouped in one area of the search space as they belong to one cluster.

\paragraph{Minimum distance $k$ for a connected graph (\kconnec)}
This metric identifies the limit $k$ (\cf~first paragraph in~Section~\ref{sec:fla:connectedness}) so that all Pareto-optimal solutions are members of one cluster~\cite{Paquete09,Liefooghe14}.
In this case, the single cluster covers the whole graph. In other words, the whole graph is connected.
Thus, \kconnec quantifies the spread of all Pareto-optimal solutions in the search space.

\begin{figure}[t!]
	\centering
	\includeSubFig{\kconnec}{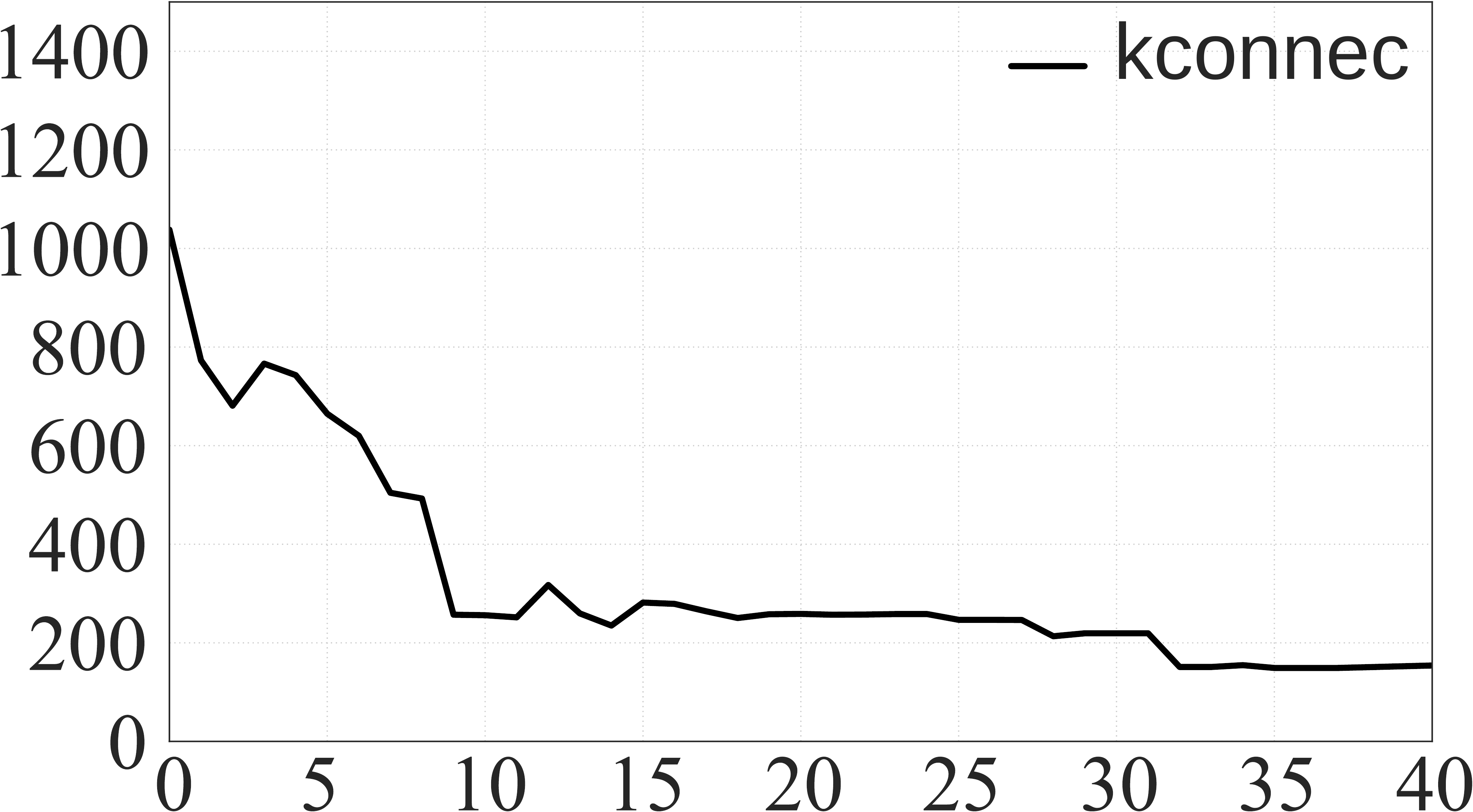}
	\caption{Minimum distance $k$ for a connected graph (\kconnec) over the 40 generations of search.}
	\label{fig:kconnec}
	\vspace{-1em}
\end{figure}

For \Sapienz, Fig.~\ref{fig:kconnec} plots \kconnec (ranging from $0$\,to\,$1400$) over the generations.
Similarly to the diameter metrics (\cf~Fig.~\ref{fig:diam}), \kconnec decreases, moderately for
\hotdeath~(from initially $\approx$700 to $\approx$600 in the first 25 generations) and
\kNineMail (from $\approx$1000 to $\approx$800 in the first 15 generations), and drastically for
\passwordmanager (from $\approx$1200 to $\approx$200 in the first 15 generations),
\munchLife (from $\approx$1000 to $\approx$200 in the first 30 generations), and
\aarddict (from $\approx$600 to $\approx$100 in the first 27 generations).
Afterwards, \kconnec stays relatively constant around its low individual level for each app.

This indicates that \textit{all} Pareto-optimal solutions are getting closer to each other in the search space as the spread of the connected graph is decreasing while this process is converging.

\paragraph{Number of Pareto-optimal solutions in the largest cluster (\lconnec)}
This metric determines the size of the largest cluster by the number of its members~\cite{Liefooghe14}, showing how many Pareto-optimal solutions are located in the most dense area of the search space. Thus, this metric counts the number of vertices of the graph that are members of the largest cluster.

\begin{figure}[b!]
	\vspace{-1em}
	\centering
	\includeSubFig{\lconnec}{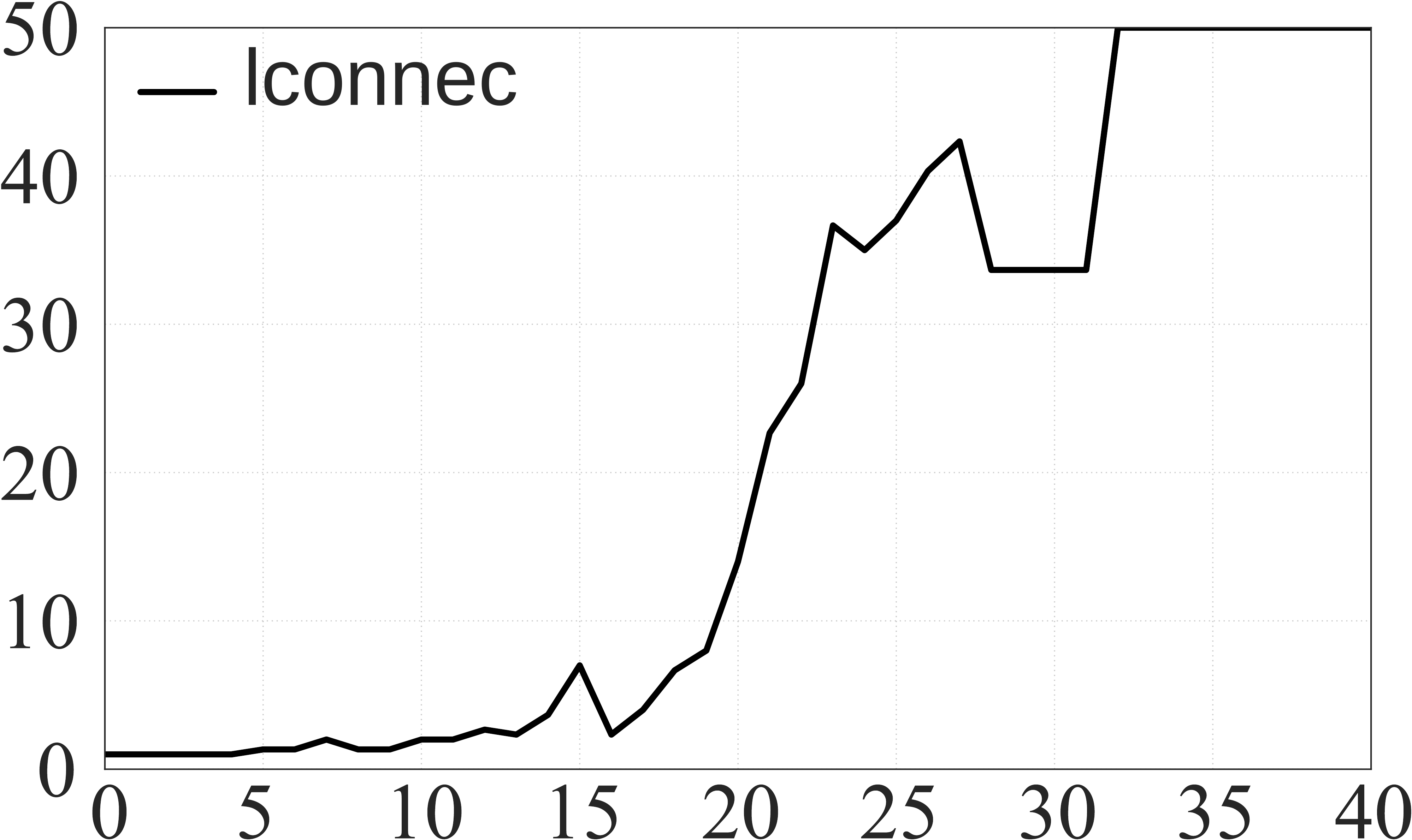}
	\caption{Number of Pareto-optimal solutions in the largest cluster (\lconnec) over the 40 generations of search.}
	\label{fig:lconnec}
	\vspace{-1em}
\end{figure}

Fig.~\ref{fig:lconnec} plots \lconnec ranging from $1$ to $50$ given the population size of $50$ (\cf~Section~\ref{sec:fla-design}) over the generations.
\lconnec increases after $15$-$30$ generations to $20$ (\aarddict and \hotdeath) or even $50$ (\munchLife) solutions. This indicates that the largest cluster is indeed large so that many Pareto-optimal solutions are grouped in \textit{one} area of the search space. In contrast, \lconnec stays always below $10$ indicating smaller largest clusters for \passwordmanager and \kNineMail than for the other apps.

\lconnec basically follows the trend of the proportion of Pareto-optimal solutions (\ppos) in Fig.~\ref{fig:ppos} since a low \ppos automatically leads to a low number of vertices in the graph (Pareto-optimal solutions) and thus, also a lower number of such solutions in the largest cluster.

\paragraph{Proportion of hypervolume covered by the largest cluster (\hvconnec)}

\begin{figure}[b!]
	\vspace{-1em}
	\centering
	\includeSubFig{\hvconnec}{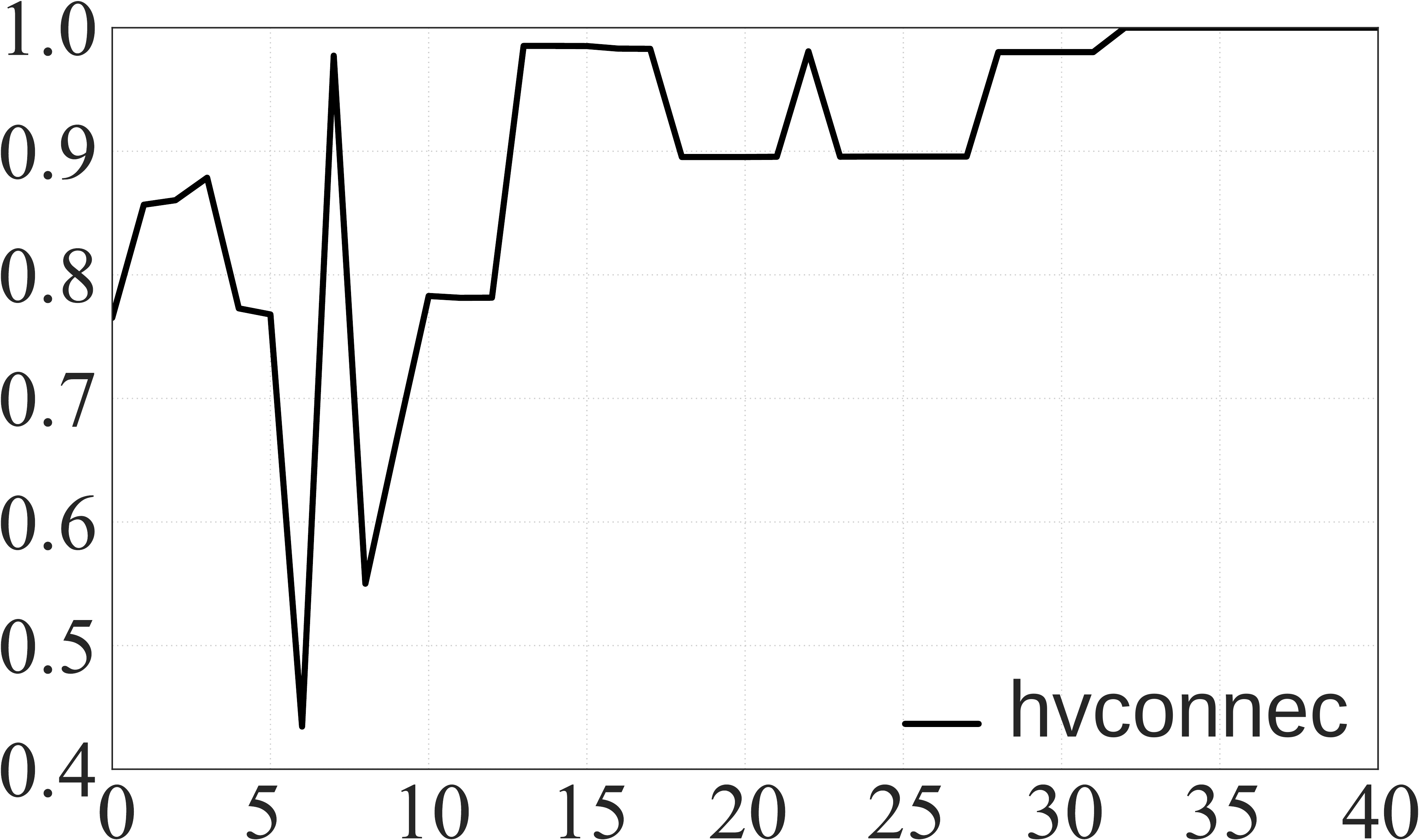}
	\caption{Proportion of hypervolume covered by the largest cluster (\hvconnec) over the 40 generations of search.}
	\label{fig:hvconnec}
\end{figure}

Besides the size of the largest cluster of the graph in terms of number of solutions (\lconnec), we compute the relative size of the largest cluster in terms of hypervolume (\hv). Thus, \hvconnec is the proportion of the overall \hv covered by the Pareto-optimal solutions in the largest cluster. It quantifies how the largest cluster in the search space dominates in the objective space and contributes to the overall \hv.
The higher \hvconnec, the more the largest cluster contributes to the \hv.

For \Sapienz (\cf~Fig.~\ref{fig:hvconnec}), \hvconnec varies a lot during the first $10$ generations, then stabilizes at a high level for all apps.
For \aarddict, \munchLife, and \passwordmanager, the largest clusters covers $100\%$ of the \hv since there is only one cluster left (\cf~\nconnec in Fig.~\ref{fig:nconnec}).
For \hotdeath, \hvconnec is close to $70\%$ indicating that there is one other cluster covering $30\%$ of the \hv (\cf~\nconnec for \hotdeath).
For \kNineMail, \hvconnec is around $90\%$ indicating that the other $2$--$3$ clusters (\cf~\nconnec for \kNineMail) cover only $10\%$ of the \hv.
This indicates that the largest cluster covers the largest proportion of the \hv, and thus contributes most to the Pareto front.

\subsection{Discussion}
\label{sec:fla-discussion}

\newcolumntype{L}{>{\raggedright\arraybackslash}m{35mm}}
\renewcommand{\arraystretch}{1}
\setlength{\tabcolsep}{4pt}

\begin{table}[]
	\centering
	\caption{Summary of the metrics and insights of the fitness landscape analysis of \Sapienz.}
	\label{tab:fla-summary}
	\resizebox*{1\textwidth}{!}{
		\tiny
		\begin{tabular}{rlp{45mm}p{45mm}} \toprule
			\cellcolor{gray!30} \# & \cellcolor{gray!30} Metric & \cellcolor{gray!30} General description & \cellcolor{gray!30} Insights for \Sapienz \\ \midrule
			\multicolumn{4}{c}{\cellcolor{gray!05} Metrics for the evolvability of the search} \\ \midrule
			1 & \ppos
			& The \textit{proportion of Pareto-optimal solutions} measures the size of the Pareto front relative to the population size. A high and especially strongly increasing \ppos may indicate a problem of dominance resistance.
			& A drastically increasing \ppos has been observed for 2/5 apps and to a smaller extent for further 2/5 apps indicating that the search of \Sapienz could suffer from \textbf{dominance resistance}. However, for 1/5 apps \ppos remained low. \\
			2 & \hv
			& The \textit{hypervolume} is the volume in the objective space covered by the Pareto-optimal solutions. An increasing \hv indicates that the search is able to find improved solutions, otherwise the \hv and search stagnate.
			& The \hv stagnates for 4/5 apps after 25 and for 1/5 apps already after five generations indicating a \textbf{stagnation of the search} in \Sapienz, potentially relatively early in the search process. \\

			\midrule
			\multicolumn{4}{c}{\cellcolor{gray!05} Metrics for the diversity of the population} \\
			\midrule
			3 & \maxdiam & The \textit{maximum population diameter} quantifies the largest distance between any two individuals and thus the maximum spread of the population in the search space.
			& During the first 25 generations, \maxdiam decreases for all apps, in 3/5 apps even drastically, denoting a \textbf{decrease of the maximum spread of the population} in the search space. \\
			4 & \avgdiam & The \textit{average population diameter} is the average of all pairwise distances between all individuals, which quantifies the average diversity of the population.
			& During the first 25 generations, \avgdiam decreases for 5/5 apps drastically denoting a \textbf{decrease of the average population diversity}. \\
			5 & \mindiam & The \textit{minimum population diameter} quantifies the smallest distance between any two individuals and thus the minimum diversity of the population.
			& The \mindiam decreases for 5/5 apps quickly reaching a value of 0, which denotes that the population contains \textbf{duplicate solutions}. \\
			6 & \reldiam & The \textit{relative population diameter} is \avgdiam divided by the largest possible distance between two individuals. It indicates the concentration of the population in the search space.
			& During the first 25 generations, \reldiam decreases for 5/5 apps indicating a \textbf{concentration/grouping of all individuals} of the population in the search space. \\

			\midrule
			\multicolumn{4}{c}{\cellcolor{gray!05} Metrics for the diversity of the Pareto-optimal solutions} \\
			\midrule
			7 & \pconnec & The \textit{proportion of Pareto-optimal solutions in clusters} indicates the degree of grouping (clustering) of Pareto-optimal solutions in the search space.
			& \pconnec increases so that for 2/5 apps most and for 3/5 apps even all \textbf{Pareto-optimal solutions are part of clusters in the search space}. \\
			8 & \nconnec & The \textit{number of clusters} counts in how many regions of the search space the Pareto-optimal solutions are clustered.
			& For 3/5 apps, there exists \textbf{one cluster containing all Pareto-optimal solutions} except of 2/5 apps with either one or two, or more than three clusters. \\
			9 & \kconnec & The \textit{minimum distance k for a connected graph} determines the maximum spread of all Pareto-optimal solutions in the search space.
			& \kconnec decreases drastically for 3/5 and moderately for 2/5 apps, denoting a \textbf{decreasing spread of Pareto-optimal solutions} in the search space. \\
			10 & \lconnec & The \textit{number of Pareto-optimal solutions in the largest cluster} quantifies the absolute size of the largest cluster and thus, how many Pareto-optimal solutions are located in the most dense area of the search space.
			& \lconnec increases for 3/5 apps denoting that the \textbf{largest cluster contains many} (20--50) \textbf{Pareto-optimal solutions}, whereas for 2/5 apps the largest cluster contains $<$10 solutions given a smaller \ppos.\\
			11 & \hvconnec & The \textit{proportion of hypervolume covered by the largest cluster} quantifies the size of the largest cluster in the objective space, that is, how much the most dense area of the search space contributes to the overall hypervolume. 			
			& For 5/5 apps, \hvconnec converges to a high value so that the \textbf{overall hypervolume is mainly (2/5) or completely (3/5) achieved by the largest cluster} making the remaining clusters less relevant in terms of fitness. \\
			
			\bottomrule
		\end{tabular}
	}
	\vspace{-1em}
\end{table}

The results characterizing the fitness landscape of \Sapienz that we presented in the previous sections reveal insights about how \Sapienz manages the search problem of generating test suites for apps. A summary of these insights gained by each of the \numberOfMetrics metrics is provided by Table~\ref{tab:fla-summary}.

Firstly, the strongly increasing proportion of Pareto-optimal solutions (\ppos) for some apps may indicate a problem of \textit{dominance resistance}, that is, the search cannot produce new solutions by crossover and mutation that dominate the current, poorly performing but locally non-dominated solutions~\cite{Purshouse+Fleming2007}. 
In other cases, \ppos remains low, that is, the search cannot find many non-dominated solutions.
Thus, \ppos may indicate a stagnation of the search, which is witnessed by the stagnation of the hypervolume (\hv) after 25 generations of search.
Consequently, the evolvability of \Sapienz as the ability to produce test suites with better fitness declines after 25 generations.

Secondly, the development of the maximum population diameter (\maxdiam) indicates a decreasing maximum spread of the population in the search space as determined by the population's \textit{two} most distant individuals.
Furthermore, the development of the average population diameter (\avgdiam) shows a decreasing diversity of \textit{all} individuals during the search.
The population even contains duplicates of individuals, which reduce the genetic variation in the population, as indicated by the minimum population diameter (\mindiam).
The development of the relative population diameter (\reldiam) witnesses this observation and indicates that the population members are concentrated in the search space~(\cf~\cite{Bachelet99}).
Consequently, the diversity of the whole population decreases during the first 25 generations, after which \Sapienz loses its ability to produce fitter test suites (evolvability).

Thirdly, we investigated how the diversity of the Pareto-optimal solutions develops as the diversity of the whole population decreases.
The development of the proportion of Pareto-optimal solutions in clusters (\pconnec) indicates a grouping of these solutions in one or more areas of the search space.
Considering the number of clusters (\nconnec), these solutions are often grouped in one or very few clusters and thus, they are located in one or very few areas of the search space.
Regardless of the actual number of clusters, the decreasing minimum distance $k$ required to form one cluster of all Pareto-optimal solutions (\kconnec) shows a declining spread of all Pareto-optimal solutions in the search space.
Moreover, considering the largest actual cluster, we notice that this cluster is often indeed large in terms of number of Pareto-optimal solutions (\lconnec), and hypervolume covered by these solutions (\hvconnec). Even if there exist multiple clusters of Pareto-optimal solutions, the largest one still contributes most to the overall hypervolume and thus, to the final Pareto front of the search. This indicates a grouping process of the Pareto-optimal solutions toward one (large) cluster in the search space that dominates in the objective space. Hence, the most-promising test suites are found by \Sapienz in one area of the search space.
Consequently, we observe a decreasing diversity of the Pareto-optimal solutions during the search.

In summary, the fitness landscape analysis of \Sapienz conducted for the selected five apps indicates a stagnation of the search (declining evolvability) while the diversity of \textit{all} solutions decreases in the search space. Thus, the loss of diversity affects the dominated and non-dominated solutions.

\section{\SapienzDiv}
\label{sec:sapienzdiv}

Given the fitness landscape analysis results summarized in Section~\ref{sec:fla-discussion}, \Sapienz suffers from a decreasing diversity of solutions in the search space over time.
It is known that the performance of genetic algorithms is influenced by diversity~\cite{Crepinsek2013,PanichellaOPL15,Morrison1998}.
A low diversity may lead the search to a local optimum that cannot be escaped easily~\cite{Crepinsek2013,Morrison1998}.
Thus, diversity is important to address dominance resistance so that the search can produce new solutions that dominate poorly performing, locally non-dominated solutions~\cite{Purshouse+Fleming2007}.
Moreover, \citet[p.\,95]{Shir09} report that promoting diversity in the search space does not hamper ``the convergence to a precise and diverse Pareto front approximation in the objective space of the original algorithm''. This statement suggests that promoting diversity in \Sapienz will not degrade the performance of \Sapienz but might rather achieve improvements.

Therefore, we extended \Sapienz to \SapienzDiv by integrating mechanisms into the search algorithm that promote the diversity of the solutions being evolved in the search space.\footnote{\SapienzDiv is available at: \url{https://github.com/thomas-vogel/sapienzdiv-ssbse19}.}	
We developed four mechanisms that extend the \Sapienz algorithm at different steps: at the initialization, before and after the variation, and at the selection. Algorithm~\ref{alg:sapienzdiv} shows the extended search algorithm of \SapienzDiv and highlights the novel mechanisms in blue.
In the following, we discuss these mechanisms.

\paragraph{Diverse initial population}
Considering Algorithm~\ref{alg:sapienzdiv}, initializing \Sapienz comprises the following steps:
Initializing the UI model, Pareto front, test reports, and the counter for the generations (lines~\ref{init:start} and~\ref{init:end}).
Booting up the Android devices or emulators to be used for executing the app under test (line~\ref{boot-device}).
Installing the Motifcore component of \Sapienz, responsible for the exploration and evaluation of tests, to the devices or emulators (line~\ref{install-sapienz}).
Statically analyzing the app under test for seeding realistic Strings to be used by test to fill text fields in the app (line~\ref{static-analysis}).\footnote{String seeding as well as the use of the so called \textit{motif genes} in \Sapienz have been deactivated in the scope of this work similar to the evaluation of \Sapienz in~\cite{Mao+2016}.}
Instrumenting (for measuring coverage) and installing the app under test to the devices or emulators.

\algnewcommand{\LineComment}[1]{\State \(\triangleright\) #1}
\definecolor{azure}{rgb}{0.0, 0.5, 1.0}
\newcommand{\CodeColor}{\textcolor{azure}}

\begin{algorithm}[t]
	\centering
	\caption{Overall algorithm of \SapienzDiv}
	\label{alg:sapienzdiv}
	\scriptsize
	\begin{algorithmic}[1]
		\Input App under test $A$, 
		crossover probability $p$, 
		mutation probability $q$,
		maximum number of generations $g_{\mathit{max}}$, 
		population size $size_{\mathit{pop}}$,
		offspring size $size_{\mathit{off}}$, 
		\CodeColor{size of the large initial population $size_{\mathit{init}}$},
		\CodeColor{diversity threshold $div_{\mathit{limit}}$},
		\CodeColor{number of diverse solutions to include $n_{\mathit{div}}$}
		\Output UI model $M$, Pareto front $PF$, test reports $C$
		\State $M \leftarrow K_{0}; PF \leftarrow \emptyset; C \leftarrow \emptyset$; \Comment{initialization} \label{init:start}
		\State \texttt{generation $g \leftarrow 0$;} \label{init:end}
		\State \texttt{boot up devices $D$;} \Comment prepare devices/emulators that will run the app \label{boot-device}
		\State \texttt{inject MOTIFCORE into $D$;} \Comment{install \Sapienz component for exploration on $D$} \label{install-sapienz}
		\State \texttt{static analysis on $A$;} \Comment{for seeding strings to be used for text fields of $A$} \label{static-analysis}
		\State \texttt{instrument and install $A$;} \Comment{app under test is instrumented and installed on $D$} \label{install-app}
		\State \CodeColor{\texttt{initialize population $P_{\mathit{init}}$ of size $size_{\mathit{init}}$;}} \Comment{\CodeColor{initialize a large initial population}}\label{initpop-start}
		\State \CodeColor{\texttt{$P = selectMostDistant(P_{\mathit{init}}, size_{\mathit{pop}})$;}} \Comment{\CodeColor{select $size_{\mathit{pop}}$ most distant individuals from $P_{\mathit{init}}$}}\label{initpop-end}
		\State \texttt{evaluate $P$ with MOTIFCORE and update $(M, PF, C)$;} \Comment{compute the fitness of these individuals}\label{eval-init-pop}
		\State \CodeColor{\texttt{$div_{\mathit{init}} = calculateDiversity(P)$;}} \Comment{\CodeColor{calculate diversity of the initial population (Eq.~\ref{eq:avgdiam})}}\label{calculate-div-init}
		\While{$g < g_{\mathit{max}}$}
		\State $g \leftarrow g + 1$;
		\State \CodeColor{$div_{\mathit{pop}} = calculateDiversity(P)$;} \Comment{\CodeColor{calculate diversity of the current population (Eq.~\ref{eq:avgdiam})}} \label{calculate-div}
		\If {\CodeColor{$div_{\mathit{pop}} \leq div_{\mathit{limit}} \times div_{\mathit{init}}$}} \Comment{\CodeColor{check decrease of diversity}} \label{check-div}
		\State \CodeColor{\texttt{$Q \leftarrow$ generate offspring of size  $size_{\mathit{off}}$;}} \Comment{\CodeColor{$\approx$ generate a new population}} \label{gen-offspring}
		\State \texttt{evaluate $Q$ with MOTIFCORE and update $(M,PF,C)$;} \Comment{compute the fitness of the offspring}\label{eval-Q-1}
		\State \CodeColor{\texttt{$P = selectMostDistant(P \cup Q, |P|)$;}} \Comment{\CodeColor{selection based on distance}} \label{selection-distance}
		\Else
		\State $Q \leftarrow wholeTestSuiteVariation(P, p, q);$ \Comment{create offspring} \label{variation}
		\State \texttt{evaluate $Q$ with MOTIFCORE and update $(M,PF,C)$;} \Comment{compute the fitness of the offspring}\label{eval-Q-2}
		\State \CodeColor{$PQ \leftarrow removeDuplicates(P \cup Q)$;} \Comment{\CodeColor{duplicate elimination}} \label{duplicate-elim}
		\State $\mathcal{F} \leftarrow sortNonDominated(PQ,|P|)$; \label{nsga2sort-1}
		\State $P' \leftarrow \emptyset$; \Comment{non-dominated individuals}
		\For{\textbf{each} front $F$ in $\mathcal{F}$}
		\If{$|P'| \geq  |P|$} break; \EndIf
		\State \texttt{$assignCrowdingDistance(F)$;}
		\For{\textbf{each} individual $f$ in $F$}
		\State $P' \leftarrow P' \cup f;$
		\EndFor
		\EndFor
		\State $P' \leftarrow sorted(P', \prec_{c});$ \label{nsga2sort-2}
		\State \CodeColor{$P \leftarrow P'[0:(size_{\mathit{pop}} - n_{\mathit{div}})]$;} \Comment{\CodeColor{take the best $(size_{\mathit{pop}} - n_{\mathit{div}})$ solutions from $P'$}} \label{takebest}
		\State \CodeColor{$P_{\mathit{div}} = selectMostDistant(PQ, n_{\mathit{div}})$;} \Comment{\CodeColor{select the $n_{\mathit{div}}$ most distant solutions from $PQ$}} \label{selectdistant}
		\State \CodeColor{$P = P \cup P_{\mathit{div}}$;} \Comment{\CodeColor{form the next population}} \label{newpopulation}
		\EndIf
		\EndWhile
		\State \textbf{return} $(M, PF, C)$;
	\end{algorithmic}
\end{algorithm}

Afterwards, \Sapienz creates the initial population.
Since the initial population may affect the results of the search~\cite{Maaranen2006}, we assume that a diverse initial population could be a better start for the exploration.
Thus, we extend the generation of the initial population $P_{\mathit{init}}$ to promote diversity. Instead of generating $|P| = size_{\mathit{pop}}$ solutions, we generate $size_{\mathit{init}}$ solutions where $size_{\mathit{init}}>size_{\mathit{pop}}$ (line~\ref{initpop-start} in Algorithm~\ref{alg:sapienzdiv}). Then, we select those $size_{\mathit{pop}}$ solutions from $P_{\mathit{init}}$ that are most distant from each other using Algorithm~\ref{alg:distance}, to form the first population~$P$ (line~\ref{initpop-end}). Subsequently, $P$ is evaluated by computing the fitness of its members (line~\ref{eval-init-pop}).

\paragraph{Adaptive diversity control}
This mechanism dynamically controls the diversity if the population members are becoming too close in the search space relative to the initial population. It further makes the \Sapienz algorithm adaptive as it uses feedback of the search to adapt the search (\cf~\cite{Crepinsek2013}).

To quantify the diversity $div_{\mathit{pop}}$ of population $P$, we use the average population diameter (\avgdiam) defined in Eq.~\ref{eq:avgdiam} (Section~\ref{sec:fla:divpop}).
At the beginning of each generation, $div_{\mathit{pop}}$ of the current population $P$ is calculated (line~\ref{calculate-div}) and compared to the diversity of the initial population $div_{\mathit{init}}$ calculated once in line~\ref{calculate-div-init}. The comparison in line~\ref{check-div} checks whether $div_{\mathit{pop}}$ has decreased to less than $div_{\mathit{limit}} \times div_{\mathit{init}}$.
For example, the condition is satisfied for the given threshold $div_{\mathit{limit}}=0.4$ if the diversity of the current population ($div_{\mathit{pop}}$) has decreased to less than $40\%$ of the diversity of the initial population ($div_{\mathit{init}}$).

In this case, the offspring $Q$ is obtained by generating new solutions using the original \Sapienz method to initialize a population (line~\ref{gen-offspring}). After evaluating the fitness of the offspring (line~\ref{eval-Q-1}), the next population is formed by selecting the $|P|$ most distant individuals from the current population $P$ and offspring $Q$ (line~\ref{selection-distance}).
In the other case, when the diversity of $P$ is at an acceptable level,
the variation operators (crossover and mutation) of \Sapienz are applied to obtain the offspring (line~\ref{variation}). After evaluating the fitness of the offspring (line~\ref{eval-Q-2}), the selection based on \nsgaTwo is applied (lines~\ref{nsga2sort-1}--\ref{nsga2sort-2}).
Thus, this mechanism promotes diversity on demand by inserting new individuals to the population, having an effect of restarting the search.

\paragraph{Duplicate elimination}
The fitness landscape analysis found duplicated test suites in the population. Eliminating duplicates is one technique to maintain diversity and improve search performance~\cite{Crepinsek2013,Ronald1998,Yuen+Chow2009}.
Thus, we remove duplicates after reproduction and before selection in the current population and offspring (line~\ref{duplicate-elim}). Duplicated test suites are identified by a distance of $0$ computed by our distance metric $dist(t_1, t_2)$ defined in Algorithm~\ref{alg:distance}.

\paragraph{Hybrid selection}
This mechanism extends the selection in \Sapienz to promote diversity in the search space. For this purpose, the selection is divided in two parts:
(1)~The non-dominated sorting of \nsgaTwo is performed as in \Sapienz (lines~\ref{nsga2sort-1}--\ref{nsga2sort-2} in Algorithm~\ref{alg:sapienzdiv}) to obtain the solutions $P'$ sorted by domination rank and crowding distance.
(2)~From $P'$, the best $(size_{\mathit{pop}} - n_{\mathit{div}})$ solutions form the next population $P$ where $size_{\mathit{pop}}$ is the size of $P$ and $n_{\mathit{div}}$ the configurable number of diverse solutions to be included in $P$ (line~\ref{takebest}). These $n_{\mathit{div}}$ diverse solutions $P_{\mathit{div}}$ are selected as the most distant solutions from the current population and offspring $PQ$ (line~\ref{selectdistant}) using our distance metric $dist(t_1, t_2)$ defined in Algorithm~\ref{alg:distance}. Finally, $P_{\mathit{div}}$ is added to the next population $P$ (line~\ref{newpopulation}).
After selection, the next evolutionary epoch~begins.

While the \nsgaTwo sorting considers the diversity of solutions in the objective space (crowding distance), the selection of \SapienzDiv also considers the diversity of solutions in the search space. Thus, the selection of \SapienzDiv is hybrid taking the objective and search space into account.

\section{Evaluation}
\label{sec:evaluation}

We evaluate \SapienzDiv in a head-to-head comparison with \Sapienz to investigate the benefits of the diversity-promoting mechanisms.
We consider \Sapienz as the baseline for our approach since \citet{Mao+2016} have already shown that \Sapienz significantly outperforms Monkey (random testing)~\cite{monkey} and Dynodroid~\cite{Machiry2013}. Thus, we do not compare \SapienzDiv against random testing. We also do not compare \SapienzDiv against other app testing approaches (\eg, Dynodroid) since we want to evaluate whether our adaptation of \Sapienz informed by the fitness landscape analysis of \Sapienz achieves any improvements over the baseline being naturally \Sapienz.
In theory, we could have selected any search-based app testing approach to analyze its fitness landscape and adapt accordingly its heuristic, in which case the selected approach would be the baseline. We selected \Sapienz as it was the state of the art at the time when we have started this work.
Thus, our evaluation is a head-to-head comparison between \SapienzDiv and \Sapienz targeting the following four research questions (RQ):
\begin{itemize}\itemsep-.4em
	\item[\textbf{RQ1}] How does the runtime overhead of \SapienzDiv compare to \Sapienz?
	\item[\textbf{RQ2}] How does the coverage achieved by \SapienzDiv compare to \Sapienz?
	\item[\textbf{RQ3}] How do the faults found by \SapienzDiv compare to \Sapienz?
	\item[\textbf{RQ4}] How does \SapienzDiv compare to \Sapienz concerning the length of their fault-revealing test sequences?
\end{itemize}

Research questions similar to RQ2, RQ3, and RQ4 have been investigated by \citet{Mao+2016} to evaluate \Sapienz. Since \SapienzDiv adds diversity-promoting mechanisms to \Sapienz, it adds a runtime overhead to \Sapienz that we investigate with RQ1.
To answer all RQs, we conduct a comprehensive empirical study that we discuss in the following.

\subsection{Novelty With Respect to Our Previous Study}

The comprehensive study presented in this article extends our study from previous work~\cite{2019-SSBSE}.
Our previous study was preliminary since we run each \Sapienz and \SapienzDiv 20 times on 10 apps while limiting the search to 10 generations because of the high execution costs of \Sapienz and \SapienzDiv. Still, the experiment for the previous study required in total almost 32 days of computation time (measured wall-clock time of the whole experiment).
For this article, we conduct a new comprehensive study by executing each \Sapienz and \SapienzDiv 30 times on 34 apps while performing the search over 40 generations. The new experiment for the comprehensive study required in total 562 days of computation time (Table~\ref{tab:experiments}).

\begin{table}[t!]
	\small
	\centering
	\caption{Overview of the new study extending our previous study~\cite{2019-SSBSE}.}
	\label{tab:experiments}
	\resizebox*{.8\textwidth}{!}{
		\begin{tabular}{lrr}
			\toprule
			& \textbf{Previous study~\cite{2019-SSBSE}} & \textbf{New study} \\
			\midrule
			\# Apps	&	10 & 34  \\
			\# Generations	& 10 & 40 \\
			\# Repetitions & 20	& 30 \\
			Total computation time (days) & 32 & 562 \\
			\bottomrule
		\end{tabular}
	}
\end{table}

For the new study, we selected the 34 apps follows.
We selected nine of the ten apps\footnote{We excluded \textit{Droidsat} as this app requests information from a website whose owner complained about the traffic caused by the tests generated and executed by \Sapienz and \SapienzDiv while conducting the experiments.} used to statistically evaluate \Sapienz~\cite{Mao+2016}. These apps have also been used in our preliminary study.
As novel apps, we further selected 21 apps from \citet{Su2017} and we choose randomly four apps from F-Droid.

Besides testing more apps with more repetitions, the major difference between the previous and new study is the number of generations that an approach searches for test suites for an app. Extending the number of generations from $10$ to $40$ is motivated by the results of the fitness landscape analysis (\cf~Section~\ref{sec:fla-sapienz}) showing that the search of \Sapienz stagnates after 25 generations. Thus, this extension allows us to investigate whether the diversity-promoting mechanisms introduced in \SapienzDiv will have an effect on the performance after the search of \Sapienz actually stagnates.

\subsection{Experimental Setup and Research Protocol}
\label{sec:evaluation:setup}

To answer the four research questions, we conducted an empirical study, in which we compare the two approaches \Sapienz and \SapienzDiv by measuring and analyzing their performance in terms of execution time, achieved coverage, triggered faults, and length of the fault-revealing test cases.
Thus, the independent variable is the testing approach and it has two levels: \Sapienz and \SapienzDiv. The dependent variables are the execution times, achieved coverage, triggered faults, and length of fault-revealing test cases.
For this purpose, we execute each approach on each of the 34 subjects (apps) whose selection was discussed in the previous section. For each approach and app, we observe and analyze the performance of the approach after 40 generations of search.\footnote{To take the overhead of \SapienzDiv into account, we will later reduce the number of generations for \SapienzDiv to enable a fair comparison with \Sapienz concerning the achieved coverage, triggered faults, and the length of fault-revealing test cases.} To enable a statistical analysis, we repeat the execution for each app and each approach 30 times.
The execution of the experiment was distributed on eight servers\footnote{For each server: 2$\times$Intel(R) Xeon(R) CPU E5-2620 @ 2.00GHz, with 64GB RAM.} where each server runs one approach to test one app at a time using ten Android emulators in parallel (Android KitKat, API~19).

For each concern (\ie, execution time, achieved coverage, triggered faults, and length of the fault-revealing test cases), we compare \Sapienz and \SapienzDiv for each individual app. Thus, we compare 30 samples of \Sapienz with 30 samples of \SapienzDiv, whereas one sample is obtained by one run of an approach on the app. The sample sets between \Sapienz and \SapienzDiv are consequently independent.
Considering such independent samples and since we cannot make any assumption about the distribution of the results, we perform a non-parametric Mann-Whitney U test~\cite{mann_test_1947,arcuri_hitchhikers_2014,DEOLIVEIRANETO2019246} for each concern to check whether the measured values of the concern achieved by both approaches differ significantly for each subject with a 95\% confidence level ($p$$<$0.05).
Moreover, we use the Vargha-Delaney effect size $\hat{A}_{12}$~\cite{Atwelve} to measure the probability that running \SapienzDiv yields better concern values than running \Sapienz (\cf~\cite{DEOLIVEIRANETO2019246}). We further characterize small, medium, and large differences between \SapienzDiv and \Sapienz with respect to a concern if $\hat{A}_{12}>$ 0.56, $\hat{A}_{12}>$ 0.64, and $\hat{A}_{12}>$ 0.71 respectively. These thresholds have been proposed by \citet{Atwelve} and used by \citet{Mao+2016} in the original evaluation of \Sapienz.

To give \Sapienz and \SapienzDiv the same conditions, both use the default \Sapienz configuration~\cite{Mao+2016} that has been also used for the fitness landscape analysis (\cf~Table~\ref{tab:sapienz-config} in Section~\ref{sec:fla-design}).
Moreover, we configured the novel parameters of \SapienzDiv as follows (Table~\ref{tab:sapienzdiv-config}):
The size of the large initial population ($size_{\mathit{init}}$) is $100$,
the diversity threshold ($div_{\mathit{limit}}$) is $0.5$, and
the number of diverse solutions to be included in the selection phase ($n_{\mathit{div}}$) is $15$.
We have been conservative in setting these parameters without any parameter tuning.
$size_{\mathit{init}}$ is just twice the size of the population ($|P|=50$), $div_{\mathit{limit}}$ requires that the diversity of a population must decrease by at least 50\% compared to the diversity of the initial population to trigger the adaptive diversity control mechanism, and $n_{\mathit{div}}$ determines that just $15$ individuals out of the $50$ population members will be selected based on diversity rather than fitness.
A less conservative parameter setting would use larger values for $size_{\mathit{init}}$ and $n_{\mathit{div}}$ and a lower value for $div_{\mathit{limit}}$.

\begin{table}[t]
	\small
	\centering
	\caption{\SapienzDiv configuration.}
	\label{tab:sapienzdiv-config}
	\resizebox*{.9\textwidth}{!}{
		\begin{tabular}{lr}
			\toprule
			\textbf{Parameter} & \textbf{Value} \\
			\midrule
			Size of the large initial population ($size_{\mathit{init}}$) & 100 \\
			Diversity threshold ($div_{\mathit{limit}}$) & 0.50 \\
			Number of diverse solutions included in the selection phase ($n_{\mathit{div}}$) & 15 \\
			\bottomrule
		\end{tabular}
	}
\end{table}

\subsection{Results}

\renewcommand{\arraystretch}{.85}
\setlength{\tabcolsep}{4pt}

In the following, we present the results of the empirical study by answering each of the four research questions.

\subsubsection{Execution Time (RQ1)}
\label{sec:evaluation:time}

To determine the runtime overhead of \SapienzDiv, we measured the execution time of each approach (\Sapienz and \SapienzDiv) to test an app over $40$ generations of search.
The corresponding results for each of the 34 apps are shown Table~\ref{tab:eval:time}.
For each app the table lists
(i)~the version of the app,
(ii)~the mean, median, and standard deviation (SD) of the execution times of the 30 runs of \Sapienz as well as \SapienzDiv,
(iii)~the p-value (printed bold if $p$$<$0.05) obtained by the Mann-Whitney U test checking significant differences between the \Sapienz and \SapienzDiv samples,
(iv)~the \Atwelve effect size comparing \SapienzDiv to \Sapienz\footnote{\label{foot:effect-size}When comparing both approaches from the perspective of \Sapienz, \ie, when switching the comparison from ``\SapienzDiv vs. \Sapienz'' to ``\Sapienz vs. \SapienzDiv'', the effect size values must be inverted ($1-\text{effect size value}$).},
and
(v)~the percentagewise overhead of \SapienzDiv with respect to \Sapienz based on the median execution times.

\begin{table}[t]
	\centering
	\caption{Execution time (minutes) of both approaches running over 40 generations on each of the 34 apps.}
	\label{tab:eval:time}
	\resizebox*{1\textwidth}{!}{\begin{tabular}{|ll|rrr|rrr|r|r|r|}
\toprule
\multirow{2}{*}{\textbf{Subject}} & \multirow{2}{*}{\textbf{Version}} & \multicolumn{3}{c|}{\Sapienz} & \multicolumn{3}{c|}{\SapienzDiv} & \multirow{2}{*}{\textbf{p-value}} & \multirow{2}{*}{\Atwelve} & \multirow{2}{*}{\textbf{Overhead}} \\ 
& & \textbf{Mean} & \textbf{Median} & \textbf{~~SD} & \textbf{Mean} & \textbf{Median} & \textbf{~~SD} & & & \\ 
\midrule
BabyCare & 1.5 & 372.00 & 368.70 & 27.66 & 470.02 & 470.71 & 57.99 & \textbf{0.000000} & 0.03 \worseAndSignificant & 27.67\% \\ 
Arity & 1.27 & 268.95 & 262.98 & 26.82 & 295.09 & 296.94 & 15.33 & \textbf{0.000001} & 0.13 \worseAndSignificant & 12.92\% \\ 
JustSit & 0.3.3 & 295.57 & 290.89 & 31.58 & 331.02 & 329.90 & 17.89 & \textbf{0.000000} & 0.09 \worseAndSignificant & 13.41\% \\ 
Hydrate & 1.5 & 302.19 & 299.17 & 11.42 & 347.42 & 347.95 & 13.25 & \textbf{0.000000} & 0.01 \worseAndSignificant & 16.31\% \\ 
FillUp & 1.7.2 & 364.09 & 363.38 & 24.61 & 455.44 & 451.96 & 39.75 & \textbf{0.000000} & 0.02 \worseAndSignificant & 24.38\% \\ 
Kanji & 1.0 & 374.08 & 366.56 & 30.38 & 465.75 & 451.12 & 50.88 & \textbf{0.000000} & 0.05 \worseAndSignificant & 23.07\% \\ 
BookWorm & 1.0.18 & 332.84 & 327.43 & 27.41 & 387.77 & 386.44 & 24.11 & \textbf{0.000000} & 0.05 \worseAndSignificant & 18.02\% \\ 
Maniana & 1.26 & 422.60 & 409.19 & 51.25 & 507.08 & 498.24 & 33.01 & \textbf{0.000000} & 0.06 \worseAndSignificant & 21.76\% \\ 
L9Droid & 0.6 & 322.38 & 326.44 & 27.23 & 406.66 & 401.75 & 31.04 & \textbf{0.000000} & 0.01 \worseAndSignificant & 23.07\% \\ 
Yaab & 1.10.1 & 379.84 & 376.92 & 38.37 & 413.00 & 397.53 & 49.94 & \textbf{0.002016} & 0.28 \worseAndSignificant & 5.47\% \\ 
Budget & 4.2 & 398.11 & 397.91 & 23.11 & 468.45 & 467.57 & 27.96 & \textbf{0.000000} & 0.02 \worseAndSignificant & 17.51\% \\ 
Campyre & 1.0.1 & 317.49 & 315.50 & 25.08 & 397.75 & 386.07 & 33.01 & \textbf{0.000000} & 0.02 \worseAndSignificant & 22.37\% \\ 
URLazy & 1.0a & 208.37 & 209.20 & 6.19 & 253.04 & 249.98 & 14.53 & \textbf{0.000000} & 0.00 \worseAndSignificant & 19.49\% \\ 
ArXiv & 2.0.22 & 377.52 & 383.23 & 52.85 & 475.32 & 475.04 & 58.22 & \textbf{0.000000} & 0.11 \worseAndSignificant & 23.96\% \\ 
Cetoolbox & 1.0 & 308.14 & 309.46 & 24.19 & 399.87 & 400.57 & 17.96 & \textbf{0.000000} & 0.00 \worseAndSignificant & 29.44\% \\ 
CurrencyConverter & 1.1 & 343.27 & 346.02 & 28.56 & 433.88 & 433.99 & 23.17 & \textbf{0.000000} & 0.01 \worseAndSignificant & 25.42\% \\ 
Charmap & 1.0.1 & 479.74 & 456.31 & 94.72 & 627.55 & 615.12 & 62.36 & \textbf{0.000000} & 0.10 \worseAndSignificant & 34.80\% \\ 
Nanoconverter & 0.8.98 & 353.04 & 350.08 & 29.39 & 445.64 & 431.29 & 48.41 & \textbf{0.000000} & 0.03 \worseAndSignificant & 23.20\% \\ 
Anarxiv & 1.0 & 286.95 & 290.71 & 31.25 & 354.66 & 373.74 & 45.77 & \textbf{0.000002} & 0.15 \worseAndSignificant & 28.56\% \\ 
Kindmind & 1.0.0\_BETA & 526.35 & 494.01 & 80.69 & 760.03 & 724.07 & 158.94 & \textbf{0.000000} & 0.06 \worseAndSignificant & 46.57\% \\ 
Urforms & 1.12 & 446.86 & 444.78 & 37.45 & 556.92 & 538.33 & 65.36 & \textbf{0.000000} & 0.04 \worseAndSignificant & 21.03\% \\ 
Homemanager & 1.0.1.8 & 338.18 & 339.40 & 18.65 & 386.95 & 385.65 & 20.16 & \textbf{0.000000} & 0.04 \worseAndSignificant & 13.63\% \\ 
Remembeer & 1.3.0 & 328.46 & 334.85 & 20.15 & 429.67 & 416.05 & 52.30 & \textbf{0.000000} & 0.00 \worseAndSignificant & 24.25\% \\ 
Pockettalk & 2.5 & 223.61 & 223.73 & 10.17 & 283.32 & 283.25 & 8.34 & \textbf{0.000000} & 0.00 \worseAndSignificant & 26.60\% \\ 
Rot13 & 1.0.2 & 331.86 & 332.27 & 24.71 & 415.20 & 410.61 & 15.56 & \textbf{0.000000} & 0.00 \worseAndSignificant & 23.58\% \\ 
DroidShows & 6.2 & 275.80 & 263.44 & 36.31 & 354.62 & 319.93 & 72.96 & \textbf{0.000001} & 0.14 \worseAndSignificant & 21.44\% \\ 
Episodes & 0.7 & 277.31 & 274.20 & 22.68 & 380.89 & 380.97 & 34.36 & \textbf{0.000000} & 0.00 \worseAndSignificant & 38.94\% \\ 
Angulo & 2.0 & 274.24 & 272.82 & 13.37 & 315.58 & 314.98 & 12.30 & \textbf{0.000000} & 0.00 \worseAndSignificant & 15.46\% \\ 
Rightsalert & 0.3a & 761.54 & 754.36 & 96.06 & 838.66 & 852.32 & 69.20 & \textbf{0.000346} & 0.24 \worseAndSignificant & 12.99\% \\ 
ApkTrack & 1.1h & 388.90 & 382.21 & 38.73 & 471.67 & 442.89 & 73.74 & \textbf{0.000000} & 0.06 \worseAndSignificant & 15.88\% \\ 
Diary & 1.0 & 264.54 & 262.43 & 12.25 & 314.07 & 313.08 & 22.57 & \textbf{0.000000} & 0.02 \worseAndSignificant & 19.30\% \\ 
rtltcp & 2.2 & 293.49 & 288.48 & 20.38 & 385.84 & 373.14 & 72.16 & \textbf{0.000000} & 0.01 \worseAndSignificant & 29.35\% \\ 
Fakedawn & 1.3 & 326.72 & 322.24 & 17.18 & 434.45 & 421.64 & 49.38 & \textbf{0.000000} & 0.00 \worseAndSignificant & 30.84\% \\ 
Klaxon & 0.27 & 250.91 & 251.29 & 9.66 & 292.45 & 292.43 & 9.94 & \textbf{0.000000} & 0.00 \worseAndSignificant & 16.37\% \\ 
\bottomrule
\end{tabular}}
\end{table}

Moreover, each \Atwelve value is followed by a triangle.
This triangle points
upwards (\betterAndSignificant/\betterAndNotSignificant) if \SapienzDiv outperforms \Sapienz,
downwards (\worseAndSignificant/\worseAndNotSignificant) if \SapienzDiv is outperformed by \Sapienz,
and to the
right (\equalAndSignificant/\equalAndNotSignificant) if \SapienzDiv performs equally to \Sapienz.
A triangle is further filled black if the corresponding results are statistically significant based on the Mann-Whitney U test with a 95\% confidence level. Otherwise, the triangles are filled white.

Concerning the results of Table~\ref{tab:eval:time}, \Sapienz significantly outperforms \SapienzDiv with large effect size on 34/34 apps for execution time.\footref{foot:effect-size}
From a practical perspective, the runtime overhead of \SapienzDiv with respect to \Sapienz ranges from 5.47\% for the app \textit{Yaab} up to 46.57\% for the app \textit{Kindmind}.
In absolute terms, for the \textit{Yaab} app the median execution times is 376.92 minutes (6.28 hours) for \Sapienz and 397.53 minutes (6.63 hours) for \SapienzDiv resulting in an overhead of 20.61 minutes for a search over 40 generations. In contrast, \textit{Kindmind} exemplifies a drastic overhead of \SapienzDiv. In this case, the median execution times is 494.01 minutes (8.23 hours) for \Sapienz and 724.07 minutes (12.07 hours) for \SapienzDiv resulting in an absolute overhead of 230.06 minutes (3.83 hours) for a search over 40 generations.
Thus, the diversity-promoting mechanisms of \SapienzDiv add a significant overhead to \Sapienz.

\begin{tcolorbox}
Answer to RQ1: Based on our evaluation, we conclude that \SapienzDiv adds a significant runtime overhead to \Sapienz that was observed for 34/34 apps and that ranges from 5.47\% to 46.57\% for individual apps. In our experiments, this resulted in an absolute overhead of \SapienzDiv ranging from around 21 minutes to 3.8 hours compared to \Sapienz when generating test suites for an individual app over 40 generations.
\end{tcolorbox}

\subsubsection{Coverage (RQ2)}

To analyze the coverage achieved by \SapienzDiv and \Sapienz, we measured the final percentagewise statement coverage achieved at the end of the search by each approach.
To make a fair coverage-based comparison between \SapienzDiv and \Sapienz, we have to take the overhead of \SapienzDiv into account (\cf~Section~\ref{sec:evaluation:time}).
Consequently, for each app we consider the coverage results of \Sapienz achieved after 40 generations of search, whereas we reduce the number of generations of search for \SapienzDiv according to the app-specific runtime overhead of \SapienzDiv.
For instance, \SapienzDiv has a runtime overhead of 27.67\% for the app \textit{BabyCare} (Table~\ref{tab:eval:time}) so that we reduce the number of generations by the same fraction (40$\times$27.67\% $=$ 11.07). Thus, we consider the coverage results of \SapienzDiv for \textit{BabyCare} after 29 generations of search. Similarly, we compute for all apps the numbers of generations considered for \SapienzDiv to enable a fair comparison.

\begin{table}[t!]
	\centering
	\caption{Final statement coverage (\%) achieved by \Sapienz after 40 generations (Gen.) and by \SapienzDiv after fewer generations (Gen.) of search taking the app-specific overhead of \SapienzDiv into account.}
	\label{tab:eval:coverage}
	\resizebox*{1\textwidth}{!}{\begin{tabular}{|l|rrrr|rrrr|r|r|}
\toprule
\multirow{2}{*}{\textbf{Subject}} & \multicolumn{4}{c|}{\Sapienz} & \multicolumn{4}{c|}{\SapienzDiv} & \multirow{2}{*}{\textbf{p-value}} & \multirow{2}{*}{\Atwelve} \\ 
& \textbf{Gen.} & \textbf{Mean} & \textbf{Median} & \textbf{~~SD} & \textbf{Gen.} & \textbf{Mean} & \textbf{Median} & \textbf{~~SD} & & \\ 
\midrule
BabyCare & 40 & 39.00 & 39.00 & 1.64 & 29 & 38.73 & 39.00 & 1.05 & 0.361919 & 0.47 \worseAndNotSignificant \\ 
Arity & 40 & 77.53 & 78.00 & 1.33 & 35 & 78.00 & 78.00 & 1.02 & 0.065616 & 0.61 \betterAndNotSignificant \\ 
JustSit & 40 & 64.53 & 64.00 & 1.36 & 35 & 64.93 & 65.00 & 1.80 & 0.237130 & 0.55 \betterAndNotSignificant \\ 
Hydrate & 40 & 47.47 & 48.00 & 0.90 & 33 & 47.87 & 48.00 & 1.11 & 0.073156 & 0.60 \betterAndNotSignificant \\ 
FillUp & 40 & 44.70 & 44.50 & 2.41 & 30 & 44.03 & 44.00 & 2.70 & 0.203496 & 0.44 \worseAndNotSignificant \\ 
Kanji & 40 & 66.47 & 66.00 & 0.51 & 31 & 66.73 & 67.00 & 0.45 & \textbf{0.018679} & 0.63 \betterAndSignificant \\ 
BookWorm & 40 & 38.57 & 39.00 & 1.50 & 33 & 40.50 & 40.50 & 2.06 & \textbf{0.000190} & 0.76 \betterAndSignificant \\ 
Maniana & 40 & 49.73 & 49.50 & 1.76 & 31 & 50.33 & 50.00 & 1.56 & 0.059615 & 0.62 \betterAndNotSignificant \\ 
L9Droid & 40 & 55.40 & 55.00 & 1.73 & 31 & 56.40 & 56.50 & 1.69 & \textbf{0.010206} & 0.67 \betterAndSignificant \\ 
Yaab & 40 & 27.83 & 28.00 & 0.38 & 38 & 28.03 & 28.00 & 0.32 & \textbf{0.017111} & 0.59 \betterAndSignificant \\ 
Budget & 40 & 50.73 & 50.50 & 1.68 & 33 & 51.40 & 51.00 & 1.43 & 0.054662 & 0.62 \betterAndNotSignificant \\ 
Campyre & 40 & 13.90 & 14.00 & 0.92 & 31 & 15.00 & 15.00 & 0.59 & \textbf{0.000004} & 0.81 \betterAndSignificant \\ 
URLazy & 40 & 32.00 & 32.00 & 0.00 & 32 & 32.00 & 32.00 & 0.00 & 1.000000 & 0.50 \equalAndNotSignificant \\ 
ArXiv & 40 & 52.97 & 53.00 & 2.62 & 30 & 53.20 & 53.00 & 2.43 & 0.362501 & 0.53 \betterAndNotSignificant \\ 
Cetoolbox & 40 & 69.73 & 70.00 & 0.52 & 28 & 69.93 & 70.00 & 0.52 & 0.074087 & 0.59 \betterAndNotSignificant \\ 
CurrencyConverter & 40 & 62.00 & 62.00 & 2.18 & 30 & 62.87 & 63.00 & 1.57 & 0.079673 & 0.60 \betterAndNotSignificant \\ 
Charmap & 40 & 73.60 & 74.00 & 0.50 & 26 & 73.57 & 74.00 & 0.50 & 0.400901 & 0.48 \worseAndNotSignificant \\ 
Nanoconverter & 40 & 39.10 & 39.00 & 1.69 & 31 & 39.53 & 40.00 & 1.31 & 0.051618 & 0.62 \betterAndNotSignificant \\ 
Anarxiv & 40 & 56.03 & 56.00 & 1.30 & 29 & 56.23 & 56.00 & 1.10 & 0.218234 & 0.56 \betterAndNotSignificant \\ 
Kindmind & 40 & 46.93 & 47.00 & 1.34 & 21 & 47.13 & 47.00 & 1.07 & 0.198322 & 0.56 \betterAndNotSignificant \\ 
Urforms & 40 & 63.13 & 63.00 & 1.80 & 32 & 63.77 & 63.00 & 1.83 & 0.100651 & 0.59 \betterAndNotSignificant \\ 
Homemanager & 40 & 49.43 & 49.00 & 1.14 & 35 & 49.73 & 50.00 & 0.83 & 0.111142 & 0.59 \betterAndNotSignificant \\ 
Remembeer & 40 & 43.50 & 43.50 & 2.79 & 30 & 45.47 & 45.00 & 1.55 & \textbf{0.000381} & 0.75 \betterAndSignificant \\ 
Pockettalk & 40 & 33.00 & 33.00 & 0.00 & 29 & 33.00 & 33.00 & 0.00 & 1.000000 & 0.50 \equalAndNotSignificant \\ 
Rot13 & 40 & 71.00 & 71.00 & 0.00 & 31 & 71.00 & 71.00 & 0.00 & 1.000000 & 0.50 \equalAndNotSignificant \\ 
DroidShows & 40 & 15.67 & 16.00 & 0.55 & 31 & 16.00 & 16.00 & 1.05 & 0.070921 & 0.59 \betterAndNotSignificant \\ 
Episodes & 40 & 38.87 & 44.50 & 13.39 & 24 & 45.33 & 45.00 & 11.60 & \textbf{0.016614} & 0.66 \betterAndSignificant \\ 
Angulo & 40 & 72.57 & 73.00 & 3.91 & 34 & 74.10 & 75.00 & 3.43 & 0.082998 & 0.60 \betterAndNotSignificant \\ 
Rightsalert & 40 & 78.33 & 78.00 & 0.48 & 35 & 78.50 & 78.50 & 0.51 & 0.098578 & 0.58 \betterAndNotSignificant \\ 
ApkTrack & 40 & 67.83 & 68.00 & 2.60 & 34 & 69.03 & 69.00 & 1.38 & \textbf{0.030447} & 0.64 \betterAndSignificant \\ 
Diary & 40 & 75.43 & 75.00 & 0.77 & 32 & 75.33 & 75.00 & 0.61 & 0.322025 & 0.47 \worseAndNotSignificant \\ 
rtltcp & 40 & 36.63 & 37.00 & 2.20 & 28 & 36.97 & 37.00 & 1.96 & 0.341988 & 0.53 \betterAndNotSignificant \\ 
Fakedawn & 40 & 54.57 & 54.00 & 1.83 & 28 & 55.00 & 54.50 & 1.91 & \textbf{0.012839} & 0.65 \betterAndSignificant \\ 
Klaxon & 40 & 39.53 & 40.00 & 0.97 & 33 & 39.63 & 40.00 & 0.67 & 0.426840 & 0.51 \betterAndNotSignificant \\ 
\bottomrule
\end{tabular}}
\end{table}

The corresponding coverage results of our study is shown in Table~\ref{tab:eval:coverage}. Similarly to Table~\ref{tab:eval:time}, this table lists for each app the mean, median, and standard deviation (SD) of the coverage results for \Sapienz and \SapienzDiv, the p-value obtained by the Mann-Whitney U test checking significant differences between the \Sapienz and \SapienzDiv coverage results, and \Atwelve effect size comparing \SapienzDiv to \Sapienz\footref{foot:effect-size}.
Moreover, the table lists the number of generations after which we consider the coverage results for the comparison---being always 40 for \Sapienz and fewer generations for \SapienzDiv.

Considering the results of Table~\ref{tab:eval:coverage}, we obtained statistically significant results for 9/34 apps.
For all of these nine apps, \SapienzDiv significantly outperforms \Sapienz with respect to coverage, in 3/9 cases with large effect size, in 3/9 cases with medium effect size, and in 3/9 cases with small effect size. In contrast, \SapienzDiv was never significantly outperformed by \Sapienz.
Concerning the statistically significant results for the nine apps from a practical perspective, \SapienzDiv was able to increase the median coverage by up to 7.14\% compared to \Sapienz (\cf \textit{Campyre} app). In absolute numbers, the increase of the median coverage is typically between 0.5 and 1.5 percentage points. Thus, the impact of the coverage improvement of \SapienzDiv is rather slightly noticeable from a practical point of view.
The remaining results for the 25/34 apps are inconclusive as they are not statistically significant.

\begin{tcolorbox}
	Answer to RQ2: Based on our evaluation of the coverage, we conclude that \SapienzDiv significantly outperforms \Sapienz on 9/34 apps, and \SapienzDiv was never significantly outperformed by \Sapienz. From a practical perspective, \SapienzDiv improves the median coverage by up to 7.14\%, in absolute numbers, however, by only up to 1.5 percentage points compared to \Sapienz.
\end{tcolorbox}

\subsubsection{Faults (RQ3)}

To compare the fault revelation capabilities of \SapienzDiv and \Sapienz, we consider a crash of the app under test as a fault.
Since the goal of testing is to find many unique crashes rather than to cause the same crash many times, the \textit{unique crashes} are of particular importance when comparing the results.
Thus, out of the total crashes we identify the unique crashes, that is, their stack traces are different from the stack traces of the other crashes of the same app caused by one approach. However, we exclude faults caused by Android (\eg, native crashes) and test harness (\eg, code~instrumentation).

\begin{table}[t!]
	\vspace{-1em}
	\centering
	\caption{Number of unique crashes revealed by \Sapienz after 40 generations (Gen.) and by \SapienzDiv after fewer generations (Gen.) of search taking the app-specific overhead of \SapienzDiv into account.}
	\label{tab:eval:crashes}
	\resizebox*{1\textwidth}{!}{\begin{tabular}{|l|rrrr|rrrr|r|r|}
\toprule
\multirow{2}{*}{\textbf{Subject}} & \multicolumn{4}{c|}{\Sapienz} & \multicolumn{4}{c|}{\SapienzDiv} & \multirow{2}{*}{\textbf{p-value}} & \multirow{2}{*}{\Atwelve} \\ 
& \textbf{Gen.} & \textbf{Mean} & \textbf{Median} & \textbf{~~SD} & \textbf{Gen.} & \textbf{Mean} & \textbf{Median} & \textbf{~~SD} & & \\ 
\midrule
BabyCare & 40 & 5.47 & 5.00 & 1.01 & 29 & 5.63 & 5.50 & 1.03 & 0.295757 & 0.54 \betterAndNotSignificant \\ 
Arity & 40 & 3.83 & 3.50 & 1.05 & 35 & 4.43 & 4.00 & 1.01 & \textbf{0.006799} & 0.68 \betterAndSignificant \\ 
JustSit & 40 & 3.73 & 3.00 & 1.23 & 35 & 3.67 & 3.00 & 0.92 & 0.429646 & 0.51 \betterAndNotSignificant \\ 
Hydrate & 40 & 0.63 & 0.00 & 0.89 & 33 & 0.80 & 1.00 & 0.85 & 0.180309 & 0.56 \betterAndNotSignificant \\ 
FillUp & 40 & 0.07 & 0.00 & 0.25 & 30 & 0.33 & 0.00 & 0.80 & 0.061647 & 0.57 \betterAndNotSignificant \\ 
Kanji & 40 & 3.47 & 3.00 & 1.04 & 31 & 3.73 & 4.00 & 0.87 & 0.154543 & 0.57 \betterAndNotSignificant \\ 
BookWorm & 40 & 2.17 & 2.00 & 1.12 & 33 & 2.47 & 2.00 & 1.41 & 0.254115 & 0.55 \betterAndNotSignificant \\ 
Maniana & 40 & 4.27 & 4.00 & 1.89 & 31 & 5.30 & 5.00 & 1.91 & \textbf{0.018722} & 0.65 \betterAndSignificant \\ 
L9Droid & 40 & 2.67 & 2.50 & 1.15 & 31 & 4.77 & 5.00 & 1.33 & \textbf{0.000000} & 0.88 \betterAndSignificant \\ 
Yaab & 40 & 1.57 & 1.00 & 0.68 & 38 & 1.87 & 1.50 & 1.17 & 0.237360 & 0.55 \betterAndNotSignificant \\ 
Budget & 40 & 6.23 & 6.00 & 0.97 & 33 & 7.20 & 7.00 & 0.92 & \textbf{0.000239} & 0.75 \betterAndSignificant \\ 
Campyre & 40 & 2.73 & 3.00 & 0.74 & 31 & 6.47 & 4.00 & 6.76 & \textbf{0.000527} & 0.74 \betterAndSignificant \\ 
URLazy & 40 & 0.07 & 0.00 & 0.25 & 32 & 0.03 & 0.00 & 0.18 & 0.285081 & 0.48 \worseAndNotSignificant \\ 
ArXiv & 40 & 2.60 & 3.00 & 0.89 & 30 & 4.90 & 4.00 & 2.17 & \textbf{0.000000} & 0.91 \betterAndSignificant \\ 
Cetoolbox & 40 & 1.63 & 2.00 & 1.07 & 28 & 2.60 & 2.50 & 1.16 & \textbf{0.001144} & 0.72 \betterAndSignificant \\ 
CurrencyConverter & 40 & 0.07 & 0.00 & 0.37 & 30 & 0.37 & 0.00 & 0.67 & \textbf{0.007549} & 0.61 \betterAndSignificant \\ 
Charmap & 40 & 0.20 & 0.00 & 0.48 & 26 & 0.20 & 0.00 & 0.41 & 0.400048 & 0.51 \betterAndNotSignificant \\ 
Nanoconverter & 40 & 2.17 & 2.00 & 1.09 & 31 & 2.83 & 3.00 & 0.75 & \textbf{0.004628} & 0.68 \betterAndSignificant \\ 
Anarxiv & 40 & 1.60 & 1.00 & 1.30 & 29 & 3.57 & 3.00 & 2.13 & \textbf{0.000137} & 0.77 \betterAndSignificant \\ 
Kindmind & 40 & 7.00 & 7.00 & 2.12 & 21 & 8.37 & 8.00 & 2.11 & \textbf{0.000494} & 0.74 \betterAndSignificant \\ 
Urforms & 40 & 5.53 & 6.00 & 1.01 & 32 & 7.00 & 7.00 & 1.93 & \textbf{0.000179} & 0.76 \betterAndSignificant \\ 
Homemanager & 40 & 11.13 & 11.00 & 1.20 & 35 & 11.60 & 11.50 & 1.16 & 0.111629 & 0.59 \betterAndNotSignificant \\ 
Remembeer & 40 & 1.67 & 1.50 & 0.76 & 30 & 2.10 & 2.00 & 0.92 & \textbf{0.032226} & 0.63 \betterAndSignificant \\ 
Pockettalk & 40 & 0.10 & 0.00 & 0.31 & 29 & 0.00 & 0.00 & 0.00 & \textbf{0.040702} & 0.45 \worseAndSignificant \\ 
Rot13 & 40 & 0.07 & 0.00 & 0.37 & 31 & 0.20 & 0.00 & 0.41 & \textbf{0.028777} & 0.58 \betterAndSignificant \\ 
DroidShows & 40 & 0.30 & 0.00 & 0.47 & 31 & 1.97 & 1.00 & 2.75 & \textbf{0.000424} & 0.73 \betterAndSignificant \\ 
Episodes & 40 & 0.13 & 0.00 & 0.43 & 24 & 0.50 & 0.00 & 0.90 & \textbf{0.015745} & 0.62 \betterAndSignificant \\ 
Angulo & 40 & 0.97 & 1.00 & 0.76 & 34 & 1.53 & 1.00 & 1.07 & \textbf{0.016776} & 0.65 \betterAndSignificant \\ 
Rightsalert & 40 & 13.90 & 14.00 & 2.04 & 35 & 15.17 & 15.00 & 1.82 & \textbf{0.003657} & 0.70 \betterAndSignificant \\ 
ApkTrack & 40 & 20.00 & 20.00 & 3.52 & 34 & 23.00 & 23.00 & 3.22 & \textbf{0.000317} & 0.76 \betterAndSignificant \\ 
Diary & 40 & 3.10 & 3.00 & 0.31 & 32 & 3.27 & 3.00 & 0.52 & 0.081530 & 0.57 \betterAndNotSignificant \\ 
rtltcp & 40 & 1.63 & 1.00 & 0.81 & 28 & 1.87 & 2.00 & 0.68 & 0.077531 & 0.60 \betterAndNotSignificant \\ 
Fakedawn & 40 & 0.17 & 0.00 & 0.46 & 28 & 0.27 & 0.00 & 0.52 & 0.172190 & 0.55 \betterAndNotSignificant \\ 
Klaxon & 40 & 0.83 & 1.00 & 0.79 & 33 & 1.43 & 1.50 & 0.90 & \textbf{0.004092} & 0.69 \betterAndSignificant \\ 
\bottomrule
\end{tabular}}
\end{table}

The corresponding results for the unique crashes revealed by \Sapienz and \SapienzDiv for each of the 34 apps are shown in Table~\ref{tab:eval:crashes} that is structured similarly to Table~\ref{tab:eval:coverage}. Again, we take the app-specific overhead of \SapienzDiv into account (\cf~Section~\ref{sec:evaluation:time}) and limit the number of generations for the search in \SapienzDiv for each app.
Overall, we obtained statistical significant results for 21/34 apps.
For 20 of these 21 apps, \SapienzDiv significantly outperforms \Sapienz with respect to the number of revealed unique crashes, in 10/20 cases with a large effect size, in 6/20 cases with a medium effect size, and in 4/20 case with a small effect size.
Concerning these results, specifically the median crashes, from a practical perspective, \SapienzDiv was able to reveal up to 3 more unique crashes of an app than \SapienzDiv (\cf median crashes for the app \textit{ApkTrack}). For the average crashes, \SapienzDiv identified up to 3.74 more crashes of an app than \Sapienz (\cf average crashes for the app \textit{Campyre}). On average across all apps, \SapienzDiv identifies around one more unique crash of an app than \Sapienz. These results for 20/34 apps illustrate the practical impact \SapienzDiv could have in revealing faults.
In contrast, \Sapienz significantly outperforms \SapienzDiv for only one app (\textit{Pockettalk}) with a very small effect size of 0.55 (or 0.45 from the perspective of \SapienzDiv). For this app, none of the 30 \SapienzDiv runs revealed any crash, whereas 3/30 \Sapienz runs revealed each a single crash.
The results for the remaining 13/34 apps are not statistically significant and therefore inconclusive.

Finally, concerning the total number of apps and the whole experiment (particularly all repetitions of runs), \Sapienz was able to reveal faults in 34/34 apps, whereas \SapienzDiv was able to reveal faults in 33/34 apps.

\begin{tcolorbox}
	Answer to RQ3: Based on our evaluation of the crashes, we conclude that \SapienzDiv significantly outperforms \Sapienz on 20/34 apps, and \Sapienz significantly outperforms \SapienzDiv on 1/34 apps. From a practical perspective, \SapienzDiv was able to reveal up to 3 and 3.74 more unique crashes of an app than \Sapienz considering the median and average crashes, respectively.
\end{tcolorbox}

\subsubsection{Length of Fault-Revealing Test Sequences (RQ4)}

To analyze the length of test sequences (cases), we consider only those test sequences that reveal faults similarly to~\cite{Mao+2016}. Moreover, to measure the length of such sequences within one run of an approach on an app, we consider the minimal length, that is, the length of the shortest of all sequences produced in this run causing the same unique crash. If more than one unique crash is found within one run, we take the average length of the corresponding minimal test sequences.

\begin{table}[t]
	\centering
	\caption{Average length of the minimal fault-revealing test sequences produced by \Sapienz after 40 generations (Gen.) and by \SapienzDiv after fewer generations (Gen.) of search taking the app-specific overhead of \SapienzDiv into account.}
	\label{tab:eval:length}
	\resizebox*{1\textwidth}{!}{\begin{tabular}{|l|rrrr|rrrr|r|r|}
\toprule
\multirow{2}{*}{\textbf{Subject}} & \multicolumn{4}{c|}{\Sapienz} & \multicolumn{4}{c|}{\SapienzDiv} & \multirow{2}{*}{\textbf{p-value}} & \multirow{2}{*}{\Atwelve} \\ 
& \textbf{Gen.} & \textbf{Mean} & \textbf{Median} & \textbf{~~SD} & \textbf{Gen.} & \textbf{Mean} & \textbf{Median} & \textbf{~~SD} & & \\ 
\midrule
BabyCare & 40 & 71.53 & 61.00 & 36.78 & 29 & 112.50 & 119.00 & 50.88 & \textbf{0.001024} & 0.27 \worseAndSignificant \\ 
Arity & 40 & 100.87 & 100.50 & 41.51 & 35 & 154.90 & 162.00 & 46.63 & \textbf{0.000027} & 0.20 \worseAndSignificant \\ 
JustSit & 40 & 216.13 & 213.00 & 61.09 & 35 & 209.83 & 211.00 & 69.31 & 0.432492 & 0.51 \betterAndNotSignificant \\ 
Hydrate & 40 & 296.69 & 301.00 & 146.72 & 33 & 321.75 & 334.00 & 105.42 & 0.421762 & 0.48 \worseAndNotSignificant \\ 
FillUp & 40 & 338.50 & 338.50 & 78.49 & 30 & 325.00 & 365.00 & 147.19 & 0.433816 & 0.50 \equalAndNotSignificant \\ 
Kanji & 40 & 74.33 & 80.50 & 50.89 & 31 & 136.17 & 135.50 & 51.86 & \textbf{0.000027} & 0.20 \worseAndSignificant \\ 
BookWorm & 40 & 101.57 & 58.00 & 88.40 & 33 & 141.67 & 126.50 & 103.73 & \textbf{0.040507} & 0.37 \worseAndSignificant \\ 
Maniana & 40 & 306.30 & 309.00 & 59.89 & 31 & 354.67 & 355.00 & 53.94 & \textbf{0.001516} & 0.28 \worseAndSignificant \\ 
L9Droid & 40 & 282.62 & 278.00 & 99.52 & 31 & 312.63 & 317.00 & 64.93 & 0.092302 & 0.40 \worseAndNotSignificant \\ 
Yaab & 40 & 135.73 & 119.50 & 63.63 & 38 & 151.00 & 116.00 & 88.03 & 0.412239 & 0.48 \worseAndNotSignificant \\ 
Budget & 40 & 243.07 & 241.00 & 44.56 & 33 & 240.57 & 237.00 & 49.69 & 0.336705 & 0.53 \betterAndNotSignificant \\ 
Campyre & 40 & 108.43 & 91.00 & 57.23 & 31 & 161.73 & 169.00 & 84.70 & \textbf{0.007978} & 0.32 \worseAndSignificant \\ 
URLazy & 40 & 254.00 & 254.00 & 315.37 & 32 & 484.00 & 484.00 & -- & 0.270146 & 0.00 \worseAndNotSignificant \\ 
ArXiv & 40 & 178.63 & 172.00 & 87.69 & 30 & 234.83 & 214.50 & 65.53 & \textbf{0.002062} & 0.28 \worseAndSignificant \\ 
Cetoolbox & 40 & 171.32 & 161.00 & 79.64 & 28 & 273.62 & 268.00 & 84.40 & \textbf{0.000051} & 0.19 \worseAndSignificant \\ 
CurrencyConverter & 40 & 271.00 & 271.00 & -- & 30 & 411.38 & 425.50 & 62.10 & 0.087622 & 0.00 \worseAndNotSignificant \\ 
Charmap & 40 & 408.80 & 434.00 & 100.18 & 26 & 400.00 & 460.50 & 153.70 & 0.463632 & 0.50 \equalAndNotSignificant \\ 
Nanoconverter & 40 & 121.81 & 124.00 & 75.55 & 31 & 210.33 & 195.00 & 84.96 & \textbf{0.000122} & 0.22 \worseAndSignificant \\ 
Anarxiv & 40 & 277.30 & 272.00 & 97.36 & 29 & 292.64 & 296.00 & 94.25 & 0.345483 & 0.47 \worseAndNotSignificant \\ 
Kindmind & 40 & 150.87 & 154.00 & 44.46 & 21 & 213.53 & 214.50 & 42.70 & \textbf{0.000002} & 0.15 \worseAndSignificant \\ 
Urforms & 40 & 219.50 & 224.00 & 54.10 & 32 & 243.17 & 232.00 & 51.81 & 0.077853 & 0.39 \worseAndNotSignificant \\ 
Homemanager & 40 & 176.10 & 172.00 & 37.21 & 35 & 194.17 & 190.50 & 42.03 & \textbf{0.042476} & 0.37 \worseAndSignificant \\ 
Remembeer & 40 & 122.60 & 102.00 & 65.16 & 30 & 173.17 & 205.00 & 99.95 & \textbf{0.031761} & 0.36 \worseAndSignificant \\ 
Pockettalk & 40 & 85.67 & 57.00 & 61.26 & 29 & -- & -- & -- & \textbf{--} & -- \phantom{\betterAndSignificant} \\ 
Rot13 & 40 & 73.00 & 73.00 & -- & 31 & 288.50 & 289.00 & 175.85 & 0.105650 & 0.00 \worseAndNotSignificant \\ 
DroidShows & 40 & 160.56 & 147.00 & 98.98 & 31 & 268.15 & 251.00 & 126.50 & \textbf{0.012572} & 0.23 \worseAndSignificant \\ 
Episodes & 40 & 381.00 & 414.00 & 91.10 & 24 & 356.40 & 430.50 & 157.00 & 0.432795 & 0.45 \worseAndNotSignificant \\ 
Angulo & 40 & 335.14 & 353.00 & 84.12 & 34 & 311.81 & 318.00 & 94.29 & 0.218887 & 0.57 \betterAndNotSignificant \\ 
Rightsalert & 40 & 203.10 & 210.00 & 33.79 & 35 & 232.30 & 237.50 & 37.21 & \textbf{0.000450} & 0.25 \worseAndSignificant \\ 
ApkTrack & 40 & 199.90 & 200.00 & 33.90 & 34 & 246.73 & 251.50 & 25.86 & \textbf{0.000000} & 0.13 \worseAndSignificant \\ 
Diary & 40 & 109.37 & 109.50 & 34.68 & 32 & 139.10 & 125.50 & 43.29 & \textbf{0.004628} & 0.30 \worseAndSignificant \\ 
rtltcp & 40 & 114.97 & 68.00 & 88.01 & 28 & 153.70 & 163.50 & 98.59 & 0.100453 & 0.40 \worseAndNotSignificant \\ 
Fakedawn & 40 & 219.75 & 170.50 & 192.32 & 28 & 273.29 & 302.00 & 162.47 & 0.318301 & 0.39 \worseAndNotSignificant \\ 
Klaxon & 40 & 315.79 & 338.00 & 122.95 & 33 & 354.52 & 360.00 & 84.12 & 0.187113 & 0.42 \worseAndNotSignificant \\ 
\bottomrule
\end{tabular}}
\end{table}

The corresponding results showing the average length of the minimal fault-revealing test sequences produced by \Sapienz and \SapienzDiv are presented in Table~\ref{tab:eval:length}. As before, we take the app-specific overhead of \SapienzDiv into account (\cf~Section~\ref{sec:evaluation:time}) and limit the number of generations for the search in \SapienzDiv for each app.
If no fault has been identified for an app by an approach in any of the 30 runs, there is no fault-revealing test sequence as denoted by `--' for the mean, median, SD, p-value, and \Atwelve columns. This is only the case for \SapienzDiv and \textit{Pockettalk}.
Moreover, if exactly one out of the 30 runs of an approach on an app has revealed faults, there is no standard deviation across the runs, which is denoted by `--' in the SD column while there exists mean and median values.

Concerning the results of Table~\ref{tab:eval:length}, we obtained statistical significant results for 16/34 apps.
For all of these 16 apps, \Sapienz significantly outperformed \SapienzDiv by producing shorter fault-revealing test sequences, in 11/16 cases with a large effect size, in 2/16 with a medium effect size, and in 3/16 cases with a small effect size.
Considering these significant results, the median length of fault-revealing test sequences produced by \SapienzDiv is at least 10.75\% (\cf \textit{Homemanager} app) and up to 118.10\% (\cf \textit{BookWorm} app) longer than the median length of corresponding sequences produced by \Sapienz. In absolute numbers, this translates to an increase of the median length of fault-revealing test sequences by 18.5 events for the \textit{Homemanager} app (median length increased from 172.0 to 190.5 events) and 68.5 events for the \textit{BookWorm} app (median length increased from 58.0 to 126.5 events).
In contrast, \SapienzDiv was not able to significantly outperform \Sapienz on any app for the length of fault-revealing test sequences.
The results for the remaining 18/34 apps are not statistically significant and therefore inconclusive.

\begin{tcolorbox}
	Answer to RQ4: Based on our evaluation of the length of minimal fault-revealing test sequences, we conclude that \Sapienz significantly outperforms \SapienzDiv on 16/34 apps by producing shorter test sequences, whereas \SapienzDiv was not able to significantly outperform \Sapienz on any app. From a practical perspective, the fault-revealing sequences of \SapienzDiv are at least 10.75\% and up to 118.10\% longer than the corresponding sequences of \Sapienz for an app.
\end{tcolorbox}

\subsection{Discussion}

The diversity-promoting mechanisms of \SapienzDiv add a significant runtime overhead to \Sapienz ranging from 5.47\% to 46.57\% for individual apps when searching over 40 generations.
The worse execution time of \SapienzDiv is caused by two factors. On the one hand, the diversity-promoting mechanisms of \SapienzDiv result in a runtime overhead, for instance, due to the more costly creation of the initial population when twice the number of individuals are generated and evaluated, and to the computation of the diversity of the population in each generation. On the other hand, \SapienzDiv produces diverse test sequences that are inherently longer than the test sequences produced by \Sapienz. Thus, executing  longer sequences on the app under test during fitness evaluation takes more time than for shorter test sequences.

Given this runtime overhead of \SapienzDiv, to fairly compare \SapienzDiv and \Sapienz, the search of \SapienzDiv has to be done with a smaller number of generations than the search of \Sapienz.
Thus, in the context of comparing the results of our empirical study, we reduced the number of generations used by the search of \SapienzDiv according to the app-specific overhead, whereas the search of \Sapienz can use all of the 40 generations. This adjustment gives both approaches the same time budget and  therefore allows us to compare fairly the achieved coverage, revealed faults, and length of minimal fault-revealing test sequences of both approaches.

\SapienzDiv significantly outperforms \Sapienz
on 9/34 apps for coverage,
on 20/34 apps for faults, and
on 0/34 apps for length.
In contrast, \Sapienz significantly outperforms \SapienzDiv
on 0/34 apps for coverage,
on 1/34 apps for faults, and
on 16/34 apps for length.
The remaining results are not statistically significant and therefore inconclusive.

Thus, \SapienzDiv achieves better or at least similar test results for coverage and faults than \Sapienz, so that preferring \SapienzDiv over \Sapienz for app testing will likely not result in disadvantages concerning fault revelation and coverage. Consequently, promoting diversity can be beneficial for generating test suites for apps.
In practical terms, \SapienzDiv revealed up to 3 and 3.74 more unique crashes of an app than \Sapienz considering the median and average crashes, respectively. On average across all apps, \SapienzDiv identified around one more unique crash of an app than \Sapienz. This illustrates the impact on fault revelation that \SapienzDiv could have in practice. In contrast, the practical impact of the improved coverage achieved by \SapienzDiv is less noticeable. \SapienzDiv was able to increase the median coverage for an individual app by up to 7.14\% compared to \Sapienz, in absolute numbers, however, by only up to 1.5 percentage points.

However, \SapienzDiv performs worse than \Sapienz in producing short fault-revealing test sequences for 16/34 apps. For the remaining 18/34 apps, we did not observe a statistically significant difference between lengths of such sequences produced by \SapienzDiv and \Sapienz. Thus, promoting diversity in \SapienzDiv tends to result in test sequences of similar or greater length. A corresponding observation has been made for diversity in unit testing resulting in longer test cases~\cite{Albunian2017}. We think that the reason for this observation is that the search can easily achieve a higher diversity between test sequences by exploring longer sequences. That is, the space of possible test sequences grows with the sequence length, and likewise the probability of generating distant sequences grows with the increase of the sequence length.
Considering the 16/34 apps, for which we obtained statistically significant results for length, the median length of fault-revealing test sequences produced by \SapienzDiv for an app is longer than the median length of corresponding sequences produced by \Sapienz, by at least 10.75\% and up to 118.10\%. Thus, the growth of the sequence length in \SapienzDiv could be acceptable but it may also result in sequences being more than double the length of the sequences produced by \Sapienz.

Consequently, when selecting \SapienzDiv over \Sapienz, a trade-off has to be made that favors potentially improved fault revelation and coverage results while accepting potentially longer test sequences that developers have to understand to debug the app and fix the fault in the app.
From a practical point of view, the improved fault revelation capabilities of \SapienzDiv are most promising since \SapienzDiv was able to find more unique crashes for most of the apps than \Sapienz (up to 3 more crashes of an app considering the median crashes). In contrast, the improved coverage achieved by \SapienzDiv seems to be less relevant in practice as the improvement is rather marginal in absolute numbers (up to 1.5 percentage points for the median coverage of an app). Thus, to adopt \SapienzDiv in practice, a trade-off needs to be primarily made between the improved fault revelation and the additional costs for developers to understand and debug longer test sequences. Such a trade-off might particularly favor \SapienzDiv over \Sapienz in cases where the increase of the sequence length is likely to be acceptable, for instance, by being only around 11\%.

\section{Threats to Validity}
\label{sec:threats-to-validity}

\paragraph{Internal validity}

A threat to the internal validity is a bias in the selection of the five apps from \cite{Mao+2016, Choudhary+2015} for the fitness landscape analysis, and of the nine apps from \citet{Mao+2016}, 21 apps from \citet{Su2017}, and four further apps we choose randomly from F-Droid for the evaluation of \SapienzDiv.
We mitigated this threat by using apps from different sources: three different research papers, while \citet{Choudhary+2015} collected their apps from four different papers, and a random sampling from F-Droid.

To reduce the threat of overfitting to the given apps, we use the default configuration of \Sapienz and \SapienzDiv without any parameter tuning.
Moreover, we conservatively set the parameters that control the diversity-promoting mechanisms in \SapienzDiv, again without any parameter tuning (\cf~Section~\ref{sec:evaluation:setup}). However, a less conservative parameter setting enforcing diversity earlier and stronger might lead to other results.

The correctness of the diversity-promoting mechanisms is a threat that we addressed by computing the fitness landscape analysis metrics with \SapienzDiv to confirm the improved diversity.
Finally, the choice of distance metric for test suites is a threat that we addressed by following recommendations from fitness landscape analysis research (\cf~Sections~\ref{sec:fla} and~\ref{sec:fl-sapienz}). Still, the results of the fitness landscape analysis and evaluation of \SapienzDiv might be different when using another metric (\eg, edit distance).

\paragraph{External validity}

As we used five different Android apps for analyzing the fitness landscape of \Sapienz and 34 different apps for evaluating \SapienzDiv out of over 2,500 apps on F-Droid and millions on Google Play, this small sample size is a potential threat to the generalizability of our findings.
To strengthen external validity, a larger number of experiments could be applied to more apps, for instance, to the 68 apps of the F-Droid benchmark by \citet{Choudhary+2015}, which, however, would imply large evaluation costs in terms of computation time.

\paragraph{Construct validity}

A threat to construct validity are the measures that we use to compare \Sapienz and \SapienzDiv.
Since the goal of testing is to find faults, we do not measure faults directly but crashes of the app caused by uncaught exceptions. We believe that this approach is justified as we consider system-level end-to-end testing, that is, tests mimic end user behavior and there should not be any uncaught exception in an app from the end user perspective.
As further measures, we consider coverage and the length of test sequences. In this context, we do not know how coverage and fault revelation are correlated, however, coverage as a secondary criterion for evaluation testing approaches is an accepted measure~\cite{Klees+2018}. Moreover, we do not know which lengths of test sequences are acceptable for developers who use such sequences to debug an app. Thus, the length of a test sequence might not be meaningful when using it as the only measure to evaluate a testing approach.
Consequently, we use the triple of crashes, coverage, and sequence length as measures, which are also the objectives of the fitness function of \Sapienz/\SapienzDiv and thus, has been used to evaluate \Sapienz~\cite{Mao+2016}.

\paragraph{Statistical and conclusion validity}

A threat to statistical and conclusion validity is our choice of statistical test (Mann-Whitney U~\cite{mann_test_1947}) and the effect size measure (Vargha-Delaney~\cite{Atwelve}).
Having data for two levels (\Sapienz and \SapienzDiv) and independent sample sets (based on 30 runs of an approach on an app), we compare both approaches for each individual app. We further cannot make any assumption about the distribution of the results. All these aspects justify our choice of the Mann-Whitney U test.
Moreover, we use the Vargha-Delaney effect size measure \Atwelve to compare \SapienzDiv and \Sapienz as recommended by~\citet{arcuri_hitchhikers_2014}. This measure is a probability estimate denoting ``how likely is it that an investigated technique \textit{X} [\eg, \textnormal{\textsc{Sapienz$^{\mathit{div}}$}}] is better than technique \textit{Y} [\eg, \textnormal{\textsc{Sapienz}}]''~\cite[p.\,261]{DEOLIVEIRANETO2019246}. A threat to conclusion validity are the probability thresholds for the effect size to distinguish large, medium, and small improvements of one technique over the other. We use the thresholds proposed by \citet{Atwelve} and used by \citet{Mao+2016} to evaluate \Sapienz.

\section{Related Work}
\label{sec:related-work}

Related work can be identified in three main areas:
approaches on test case generation for apps,
approaches on diversity in search-based software testing, and
fitness landscape analysis for software engineering problems.

\paragraph{Test case generation for mobile applications}
Strategies for generating test cases can be grouped into approaches using random, model- and search-based, or systematic exploration strategies for the generation.

Random strategies such as Monkey~\cite{monkey} implement UI-guided test input generators where events on the GUI are selected randomly.
Dynodroid~\cite{Machiry2013} extends the random selection using weights and frequencies of events.
Model-and search-based strategies such as COBWEB~\cite{JabbarvandLM19}, SwiftHand~\cite{ChoiNS13}, EHBDroid~\cite{SongQH17}, PUMA~\cite{Hao2014}, EvoDroid~\cite{Mahmood2014}, DroidBot~\cite{Li2017}, MobiGUITAR~\cite{Amalfitano2015}, juGULAR~\cite{AmalfitanoRASF19}, ABE~\cite{GuSMC0YZLS19}, Humanoid~\cite{LiYGC2019}, \Sapienz~\cite{Mao+2016}, Stoat~\cite{Su2017}, and MATE~\cite{Sell+2019} apply model-based testing with dedicated search strategies to mobile applications. The type of model and search strategy is different in each approach and can range from simple to sophisticated solutions.

For instance, COBWEB~\cite{JabbarvandLM19} extracts a graph-based energy model which is then used in combination with a genetic algorithm to construct test cases to reveal so called energy bugs. Similar evolutionary exploration strategies are used by EvoDroid~\cite{Mahmood2014} and Sapienz~\cite{Mao+2016}. SwiftHand~\cite{ChoiNS13} on the other hand uses approximate learning to learn the app's GUI model during testing. The learned GUI model enables generation of inputs that visit unexplored states of the app. Alternatively, juGULAR~\cite{AmalfitanoRASF19} uses machine learning as an automatic GUI model exploration engine.
Thus, systematic exploration strategies apply full-scale symbolic execution~\cite{Mirzaei2012} to evolutionary algorithms.
All of these approaches do not explicitly manage diversity, except of Stoat~\cite{Su2017} that encodes the diversity of sequences into the objective function. 

\paragraph{Diversity in search-based software testing}

The diversity of solutions has been researched in search-based software testing, mostly in the area of test case selection and generation.
For test case selection, Panichella~\etal~\cite{PanichellaOPL15} showed that multi-objective genetic algorithms (GAs) can be significantly improved by diversifying the solutions in the search process. They propose several diversity-preserving techniques for the main loop of \nsgaTwo. The resulting algorithm is called DIV-GA (DIVersity based GA). The conducted study showed that DIV-GA outperforms both \nsgaTwo and vNSGA-II~\cite{Yoo2007} that are considered state of the art for multi-objective optimization.

For test case generation, there are approaches that consider feature diversity \cite{Biagiola0RT19,FeldtP2017,Albunian2017} in their search procedures. The closest to our approach is the work of \citet{Biagiola0RT19} that applies diversity-based test case generation to web applications. The results show that diversity-enabled search-based testing approaches achieve a higher state coverage, code coverage, and fault detection rate than random test case generation approaches.
The work of \citet{FeldtP2017} experiments with ten different search strategies that are build on top of G{\"O}DELTEST~\cite{FeldtP13} which is a search-based data generation framework. The results show that spread-aware random sampling such as latin hypercube sampling is very effective in creating diverse test cases. Additionally, the study of Albunian~\cite{Albunian2017} matches our observation that promoting diversity during the test case generation process will increase the length of tests without improving considerably the coverage.

Current approaches for test case selection and generation witness that diversity promotion is crucial although its realization ``requires some care''~\cite[p.\,782]{Purshouse+Fleming2007}. Finally, the suitability of metrics for diversity of a test suite~\cite{ShiCFFX16} and the visualization of test diversity~\cite{NetoFEN2018} are still open research topics.

\paragraph{Fitness landscape analysis for software engineering problems}

Several approaches have analyzed fitness landscapes of various search-based software engineering problems.
For search-based testing, for instance, \citet{Waeselynck+2007} investigate the ruggedness (local structure) and size of the search space to configure a simulated annealing algorithm for test generation. The size of the search space is quantified by a diameter metric, whereas we use several diameter metrics to measure the distance between individuals of a population.
\citet{Lefticaru+Ipate2008} investigate the local structure and size of the search space similarly to \citet{Waeselynck+2007} and extend their analysis with a problem hardness measure (fitness distance correlation) that, however, requires knowledge about the global optimum, which is, for instance, not known for the problem of testing apps. Based on their analysis, various fitness functions are evaluated for specification-based testing.

More recently, \citet{Aleti16} have analyzed the evolvability and local structure of the landscape for the problem of generating whole test suites with EvoSuite.  The fitness landscape analysis used three metrics, namely population information content, negative slope coefficient, and change rate, to mainly assess solvability. 
Based on their analysis, they conclude that ``the search space is rather poor in gradients to local optima''~\cite[p.\,619]{Aleti16} and it is either unimodal or has few modes with many plateaus.
In the same direction, \citet{Nasser+2020} further analyze the local structure in terms of ruggedness and neutrality of the landscape for generating tests with the MOSA (Many-Objective Sorting Algorithm) version of EvoSuite. Similarly to \citet{Aleti16}, their analysis showed that the landscape is dominated by plateaus. Our two metrics for evolvability \ppos and \hv confirm this observation. Additionally, our fitness landscape analysis complements the studies of \citet{Aleti16} and \citet{Nasser+2020} with a detailed analysis of the diversity of the population and the diversity of the pareto-optimal solutions, which shows a drastically decreasing diversity over time.

Besides search-based testing, a fitness landscape analysis has been conducted for other search-based software engineering problems. For instance, \citet{Aleti+Moser2015} tackle the problem of optimizing software architectures while focusing on an analysis of the local structure (ruggedness).
Another line of research analyzes the fitness landscape to identify whether the landscape is elementary or not~\cite{Lu+2010}, or to construct elementary landscapes~\cite{Chicano+2011}. Such landscapes are a special class of fitness landscapes that could inform the development of suitable heuristics, for instance, for the next release problem~\cite{Lu+2010} or test suite minimization~\cite{Chicano+2011}.

In contrast to our work, none of these approaches address the problem of generating test suites for mobile apps or focus on the global topology, that is, how solutions and fitnesses are spread in the search space in terms of diversity of dominated and non-dominated solutions being evolved. This makes our previous work~\cite{2019-SSBSE} extended in this article novel.

\section{Conclusions and Future Work}
\label{sec:conclusion}

In this article, we reported on our descriptive study analyzing the fitness landscape of \Sapienz indicating a lack of diversity of test suites being evolved during the search. Therefore, we proposed \SapienzDiv that integrates four mechanisms to promote diversity.
The evaluation shows that \SapienzDiv achieves better or at least similar test results in terms of achieved coverage and revealed faults than \Sapienz, but it generates fault-revealing test sequences of similar or greater length than \Sapienz. Thus, preferring \SapienzDiv over \Sapienz for app testing could be advantageous concerning fault revelation and coverage, and disadvantageous concerning the length of test sequences that developers have to understand to debug the app and fix the fault in the app.
Thus, the understanding of the search problem obtained by the fitness landscape analysis helped us to find a more suitable configuration of \Sapienz without trial-and-error experiments.
Particularly, we think that the costs of performing the fitness landscape analysis with five apps and five repetitions would be lower than performing trial-and-error experiments with different \Sapienz configurations to gain the same insights.

This illustrates the general benefits a fitness landscape analysis could have for search-based software engineering. By investigating the fitness landscape, characteristics and particularly difficulties of a search problem (\eg, a decreasing diversity of solutions being evolved~\cite{2019-SSBSE}, or many plateaus in the landscape~\cite{Aleti16,Nasser+2020}) could be identified. Such analytical results enable a better understanding of the search problem and search algorithm, whose behavior is otherwise a black box~\cite{Culberson1998}. Moreover, the improved understanding could be leveraged to develop more suitable algorithms (\eg, in terms of operators or parameter configurations) for the given problem, especially algorithms that can cope with the identified difficulties. Without a fitness landscape analysis, improved algorithms are typically identified by empirically trying out variants of algorithms, which is a trial-and-error and costly process. With the gained understanding from a fitness landscape analysis, such a process should be avoided and the development of algorithms should be systemized.

Practitioners such as software testers would benefit from a fitness landscape analysis as it informs them on how to improve or at least appropriately configure search-based testing tools (\eg, considering the many configuration parameters of state-of-the-art heuristics) for their problem at hand.
However, in practice and research the challenge arises of how to conduct a fitness landscape analysis because many aspects aspects of fitness landscape could be analyzed (\eg, the global topology, local structure, and evolvability) with many metrics~\cite{Malan+Engelbrecht2013,Pitzer+Affenzeller2012}.
This calls for developing guidelines for software engineering researchers and practitioners of how to conduct fitness landscape analyses and for integrating corresponding mechanisms into search-based tools.

Similar to the effectiveness mapping of automated test suite generation techniques by \citet{OliveiraAGS18}, we plan to analyze the apps on which \Sapienz and \SapienzDiv perform differently to identify and map app characteristics for which a high diversity of test suites is beneficial for a search. Such characteristics could be the criteria to select \SapienzDiv---when diversity is a concern---at the costs of longer execution times, otherwise \Sapienz as the more time efficient approach would be chosen.
Finally, we plan to investigate other parameter settings controlling the novel \SapienzDiv features to enforce diversity earlier and stronger.

\vspace{1em}\noindent
{\footnotesize 
\textbf{Acknowledgments}
This work has been developed in the \textit{FLASH} project (GR 3634/6-1) funded by the German Science~Foundation (DFG) and has been partially supported by the 2018 Facebook Testing and Verification research award. We thank the Hasso Plattner Institute for providing us access to the HPI Future SOC computing infrastructure to conduct the experiments. \par
}

\setlength{\bibsep}{2.5pt plus 0.3ex}
\bibliographystyle{elsarticle-num-names}
\bibliography{ssbse19,testcasegeneration,searchspaceclassification,diversity_ssbse,testingmobile}
\end{document}